\documentclass[aps,physrev,preprint,groupedaddress]{revtex4-2}

\usepackage{graphicx}
\usepackage{subcaption}
\usepackage{xcolor}
\usepackage{cancel}
\usepackage{multirow}
\usepackage{newtxtext}
\usepackage{newtxmath}
\usepackage{lineno}
\usepackage{float}

\begin{document}


\title{Control of deterministic breakdown to turbulence of hypersonic boundary layer with spanwise non-uniform surface temperature}

\author{Luca Boscagli}
\email[]{l.boscagli@imperial.ac.uk}
\affiliation{Department of Aeronautics, Imperial College London, London, London, UK}
\author{Georgios Rigas}
\affiliation{Department of Aeronautics, Imperial College London, London, London, UK}
\author{Olaf Marxen}
\affiliation{School of Mechanical Engineering Sciences, University of Surrey, Guildford, UK}
\author{Paul J. K. Bruce}
\affiliation{Department of Aeronautics, Imperial College London, London, London, UK}

\begin{abstract}
Direct Numerical Simulation (DNS) of a Mach 6 boundary layer over a flat plate is performed to assess the effect of spanwise non-uniform surface temperature on breakdown to turbulence under deterministic forcing. The streamwise location of laminar to turbulent transition in hypersonic boundary layers has a significant influence on viscous drag and aerodynamic heating of external surfaces of hypersonic vehicles. Previous work investigated the stabilization of hypersonic boundary layers by optimally growing streaks. More recently, DNS for a hypersonic boundary layer showed that it is possible to generate streaks through a spanwise non-uniform surface temperature distribution. The laminar computations showed the control method can stabilize the second Mack mode and it is robust across a range of Mach numbers and wall temperature ratios. In this work, two scenarios are investigated where two-dimensional (second Mack mode) and oblique (first Mack mode) disturbances dominate the initial linear stage of transition. It is found that weak control streaks with amplitude below 5\% of the freestream velocity can reduce high-frequency shear-stress due to the second Mack mode by approximately 30\% relative to the uncontrolled configuration, and delay transition. For first Mack mode dominated breakdown, the control streaks have no effect on transition location, but the peak amplitude of the spanwise-integrated wall heat flux is reduced. For the first and second Mack mode-dominated scenarios, the mean and high-frequency peak heat transfer are reduced approximately by 15\% and 34\%, respectively. The dominant mechanisms are identified and attributed to the pressure work contribution to turbulent kinetic energy and the second Mack mode dilatation work. 
\end{abstract}


\maketitle

\section{Introduction}
The development of aerospace technologies that travel with a flight Mach number ($M_{\infty}$) well above sonic is challenged by complex aerothermodynamic behaviors. This flight regime is typically referred to as hypersonic. Boundary layer instability and transition can significantly affect the flight envelope and operability limits of hypersonic vehicles \citep{Lin2008}. For laminar hypersonic boundary layers, an important non dimensional parameter is the relative Mach number $\overline{M}$, which is defined based on the velocity of the flow ($u$) relative to the phase speed ($c_{ph}$) of the hydrodynamic instability within the boundary layer. When $\overline{M}^2>1$, the compressible counterpart of Rayleigh's equation admits multiple wave-like solutions, also referred to as higher Mack modes \citep{Mack1975}. For a flight Mach number between 4 and 6, a two-dimensional instability mode (second Mack mode) plays an important role within the initial linear transition stage. However, the route to turbulence in hypersonic boundary layers at flight relevant conditions is poorly understood \citep{Franko2013}, and this challenges the development of effective control strategies.

Building upon experimental \citep{Fransson2006} and numerical \citep{Cossu2002,Bagheri2007,Schlatter2010,Shahinfar2012} studies for low-speed (incompressible) boundary layers,  \citet{Paredes2016} and \citet{Ren2016} showed that finite amplitude streaks can be used to achieve successful stabilization of high-speed (compressible) boundary layers. For a Mach 2 boundary layer over an adiabatic flat plate, \citet{Sharma2019} and \citet{Kneer2022} conducted a set of parametric DNS studies and showed that streaks generated by a blowing and suction strip can successfully delay first mode oblique breakdown to turbulence. For a similar configuration, \citet{Celep2022} showed that uniform wall heating can reduce the useful range of control-streak amplitude that can successfully delay transition. For $M_{\infty}=4.5$, \citet{Zhou2023} showed that second mode oblique breakdown can also be successfully delayed through finite amplitude streaks.

Passive control of hypersonic boundary layer transition has been experimentally and numerically attempted through the use of roughness elements \cite{Marxen2010,Fong2015,Taylor2016} or vortex generators \cite{Paredes2019}. Recent computational \cite{Ozawa2025} and experimental \cite{Ozawa2025_SCITECH, Ozawa2026_SCITECH} studies showed that for a flat plate configuration it is possible to generate streaks within the boundary layer through a spanwise non-uniform wall temperature distribution. This can be passively attained through the use of alternate stripes of materials with different thermal properties, and by exploiting the high heat flux characteristics of the hypersonic regime. Effective stabilization of the second Mack mode through the novel control method can be attained using streaks with a wavelength that is approximately 8-10 times the local boundary layer thickness \citep{Boscagli2025_SCITECH}. \citet{Boscagli2026_JFM} showed that the control method is robust at flight conditions and the stabilization of the second Mack mode is driven by the base flow modification due to the streaks. This non-intrusive, passive flow control technique has the potential to increase the aero-thermal-structural efficiency of hypersonic vehicles. However, there is a need to determine its effectiveness on laminar to turbulence transition delay.

Several studies attempted to establish causality between the steady, control streaks and boundary layer transition delay. Primary interaction mechanisms are attributed to Mean Flow Deformation (MFD), $(f,k)=(0,0)$ \citep{Cossu2002,Wassermann2002}, and/or three-dimensional, steady deformation, $(f,k)=(0,k_{\neq 0})$, of the boundary layer \citep{Paredes2019,Sharma2019}. The dominant mechanism is strongly influenced by the specific transition to turbulence scenario, amplitude ($As_u$) and spanwise wavelength ($\lambda_{z,s}$) of the streaks, as well as operating conditions and configurations (table \ref{tab:review_streaks}). For a low-speed (incompressible) boundary layer over a flat plate, \citet{Bagheri2007} using the non-linear Parabolized Stability Equations (PSE) showed that the beneficial effect of finite amplitude streaks on MFD leads to a stabilization of the Tollmien-Schlichting waves. Similarly, for hypersonic (Mach 4.5 and 6) conditions and using non-linear PSE, \citet{Ren2016} showed that control-streaks with $As_{u}$ below 5\% of the freestream speed ($\tilde{u}_{\infty}$) can stabilize both first and second Mack modes through a modification of the MFD, which leads to a beneficial reduction of the momentum thickness of the boundary layer. For a Mach 5.3 cone configuration and for $As_{u}=0.1\tilde{u}_{\infty}$, \citet{Paredes2019} also showed a significant reduction of the growth of the second Mack mode planar waves via the control-streaks, but the stabilization mechanism was dominated by the three-dimensional steady deformation of the boundary layer due to the control streaks, rather than by the modification to the MFD. For the same case study \citep{Paredes2019}, when the control-streaks amplitude is increased to $As_{u}=0.2\tilde{u}_{\infty}$, the stable oblique, first Mack mode waves grow in amplitude and lead via triadic interaction \citep{Paredes2017} to streak instability that may dominate the transition process. For supersonic (Mach 2 and 3) configurations, computational studies using non-linear PSE \citep{Paredes2017} and Direct Numerical Simulations \citep{Sharma2019} showed that sufficiently closely spaced control-streaks are able to suppress oblique waves and delay transition to turbulence. \citet{Sharma2019} showed that most effective control-streaks had amplitudes approximately in the range of 10 to 20\% of the freestream streamwise momentum ($\tilde{\rho u}_{\infty}$), and wavelength $1/5^{th}$ to $1/4^{th}$ of the fundamental oblique mode . For a wavelength of the control streaks close to the value for optimal transient growth ($\lambda_{z,opt}$), the three-dimensional steady deformation of the streaks becomes effective to suppress the growth of the oblique waves.

\begin{table}
\caption{Summary of some previous findings on effective control of boundary layer stability and transition with streaks. $\lambda_{z,0}$ and $\lambda_{z,opt}$ indicate the fundamental spanwise wavelengths of the supported instability and optimally growing streaks, respectively.}
\label{tab:review_streaks}
\begin{ruledtabular}
\begin{tabular}{llllll}
\multicolumn{1}{c}{Studies} &
  \multicolumn{1}{c}{\begin{tabular}[c]{@{}c@{}}Flow regime, \\  configuration\end{tabular}} &
  \multicolumn{1}{c}{\begin{tabular}[c]{@{}c@{}}Investigation \\ method\end{tabular}} &  
  \multicolumn{1}{c}{\begin{tabular}[c]{@{}c@{}}Instability /\\ transition scenario\end{tabular}} &
  \multicolumn{1}{c}{\begin{tabular}[c]{@{}c@{}}$As_u$, \\ $\lambda_{z,s}$\end{tabular}} &
  \multicolumn{1}{c}{\begin{tabular}[c]{@{}c@{}}Dominant\\ control \\ mechanism\end{tabular}} \\ 
\multicolumn{1}{c}{\citet{Cossu2002}} &
  \multicolumn{1}{c}{\begin{tabular}[c]{@{}c@{}}Incompress., \\ Blasius profile \end{tabular}} &
  \multicolumn{1}{c}{\begin{tabular}[c]{@{}c@{}}DNS\end{tabular}} &
  \multicolumn{1}{c}{\begin{tabular}[c]{@{}c@{}}Planar TS \\ waves\end{tabular}} &
  \multicolumn{1}{c}{\begin{tabular}[c]{@{}c@{}}$[0.14,0.26]\tilde{u}_{\infty}$, \\ $\lambda_{z,opt}$\end{tabular}} &
  \multicolumn{1}{c}{MFD} \\   
\multicolumn{1}{c}{\citet{Bagheri2007}} &
  \multicolumn{1}{c}{\begin{tabular}[c]{@{}c@{}}Incompress., \\ flat plate\end{tabular}} &
  \multicolumn{1}{c}{\begin{tabular}[c]{@{}c@{}}Non-linear \\ PSE\end{tabular}} &
  \multicolumn{1}{c}{\begin{tabular}[c]{@{}c@{}}Planar TS \\ waves\end{tabular}} &
  \multicolumn{1}{c}{\begin{tabular}[c]{@{}c@{}}$[0.15,0.20]\tilde{u}_{\infty}$, \\ $[0.69,1]\lambda_{z,opt}$\end{tabular}} &
  \multicolumn{1}{c}{MFD} \\
\multicolumn{1}{c}{\citet{Bagheri2007}} &
  \multicolumn{1}{c}{\begin{tabular}[c]{@{}c@{}}Incompress., \\ flat plate\end{tabular}} &
  \multicolumn{1}{c}{\begin{tabular}[c]{@{}c@{}}Non-linear \\ PSE\end{tabular}} &
  \multicolumn{1}{c}{\begin{tabular}[c]{@{}c@{}}Oblique \\ waves\end{tabular}} &
  \multicolumn{1}{c}{\begin{tabular}[c]{@{}c@{}}$0.1\tilde{u}_{\infty}$, \\ $[0.2,0.25]\lambda_{z,0}$\end{tabular}} &
  \multicolumn{1}{c}{MFD} \\
\multicolumn{1}{c}{\citet{Ren2016}} &
  \multicolumn{1}{c}{\begin{tabular}[c]{@{}c@{}}Mach 4.5, 6, \\ flat plate\end{tabular}} &
  \multicolumn{1}{c}{\begin{tabular}[c]{@{}c@{}}Non-linear \\ PSE\end{tabular}} &
  \multicolumn{1}{c}{\begin{tabular}[c]{@{}c@{}}First and second \\ Mack modes\end{tabular}} &
  \multicolumn{1}{c}{\begin{tabular}[c]{@{}c@{}}$\leq 0.05\tilde{u}_{\infty}$, \\ $\approx 2\lambda_{z,opt}$\end{tabular}} &
  \multicolumn{1}{c}{MFD} \\
\multicolumn{1}{c}{\citet{Paredes2017}} &
  \multicolumn{1}{c}{\begin{tabular}[c]{@{}c@{}}Mach 3, \\ cone\end{tabular}} &
  \multicolumn{1}{c}{\begin{tabular}[c]{@{}c@{}}Non-linear \\ PSE\end{tabular}} &
  \multicolumn{1}{c}{\begin{tabular}[c]{@{}c@{}}First \\ Mack mode\end{tabular}} &
  \multicolumn{1}{c}{\begin{tabular}[c]{@{}c@{}}$[0.05,0.2]\tilde{u}_{\infty}$, \\ $[0.25,0.4]\lambda_{z,0}$\end{tabular}} &
  \multicolumn{1}{c}{MFD} \\
\multicolumn{1}{c}{\citet{Paredes2019}} &
  \multicolumn{1}{c}{\begin{tabular}[c]{@{}c@{}}Mach 5.3, \\ cone\end{tabular}} &
  \multicolumn{1}{c}{\begin{tabular}[c]{@{}c@{}}Non-linear \\ PSE\end{tabular}} &
  \multicolumn{1}{c}{\begin{tabular}[c]{@{}c@{}}Second \\ Mack mode\end{tabular}} &
  \multicolumn{1}{c}{\begin{tabular}[c]{@{}c@{}}$0.1\tilde{u}_{\infty}$, \\ $[0.7,1]\lambda_{z,opt}$\end{tabular}} &
  \multicolumn{1}{c}{$(0,k_{\neq 0})$} \\ 
\multicolumn{1}{c}{\citet{Sharma2019}} &
  \multicolumn{1}{c}{\begin{tabular}[c]{@{}c@{}}Mach 2, \\ flat plate\end{tabular}} &
  \multicolumn{1}{c}{DNS} &
  \multicolumn{1}{c}{\begin{tabular}[c]{@{}c@{}}Oblique \\ (first Mack mode) \\ breakdown\end{tabular}} &
  \multicolumn{1}{c}{\begin{tabular}[c]{@{}c@{}}$[0.1,0.20]\tilde{\rho u}_{\infty}$, \\ $[0.2,0.25]\lambda_{z,0}$\end{tabular}} &
  \multicolumn{1}{c}{\begin{tabular}[c]{@{}c@{}}MFD and \\ $(0,k_{\neq 0})$\end{tabular}} 
\end{tabular}
\end{ruledtabular}
\end{table}
Overall, the aim of this work is to assess, via Direct Numerical Simulations (DNS), the effect of streaks generated through non-uniform surface temperature distributions on boundary layer stability and transition to turbulence. In contrast to \citet{Boscagli2026_JFM}, which focused on laminar-regime stabilization of the second Mack mode, the present study addresses the transition process itself, including nonlinear breakdown and turbulence onset under conditions where both first- and second-mode mechanisms can be dominant, and the effect of the control-streaks on heat-transfer peaks.

Section \ref{sec:methodology} presents the case study, computational domain and numerical methods used. Results and discussion are presented in sections \ref{sec:transition_SMF} and \ref{sec:transition_FMO}, where the effect of streaks on laminar to turbulent transition under deterministic forcing is determined and quantified. The effect of control-streaks on the late stage of transition to turbulence and heat-transfer peaks is evaluated in section \ref{sec:heat_transfer_analysis}. Finally, a summary of the impact and contribution of this work is presented in section \ref{sec:conclusions}.

\section{Methodology}\label{sec:methodology}
The effect of finite amplitude, control-streaks on laminar to turbulent transition for a high-speed boundary layer over a flat plate with zero pressure gradient is investigated by means of Direct Numerical Simulations (DNS). In the sections below a description of the numerical methods and notation, the formulation of the wall boundary conditions and the data analysis methodology used is provided.

\subsection{Direct numerical simulations}\label{sec:DNS}
\subsubsection{Governing equations and numerical method}
The compressible, time-dependent formulation of the Navier-Stokes equations is numerically solved for a calorically perfect gas (air). The governing equations, specifically conservation of mass, balance of momentum and energy conservation are expressed in non-dimensional form as:

\begin{equation}
\frac{\partial \rho}{\partial t} + \frac{\partial}{\partial x_j} \left(\rho u_j \right) = 0 \, ,
\end{equation}

\begin{equation}
\frac{\partial \rho u_i}{\partial t} + \frac{\partial}{\partial x_j} \left(\rho u_i u_j + p\delta_{ij} \right) = \frac{\partial \sigma_{ij}}{\partial x_j} \, ,
\end{equation}

\begin{equation}
\frac{\partial E}{\partial t} + \frac{\partial}{\partial x_j} \left[ \left(E + p \right)u_j \right] = -\frac{\partial q_j}{\partial x_j} + \frac{\partial}{\partial x_k} \left(u_j \sigma_{jk} \right) \, .
\end{equation}
Einstein notation is used in the preceding equations, and $\sigma_{ij}$, $E$ and $q_j$ are the viscous stress tensor, the total energy per unit volume, and the heat flux vector, respectively, and these are expressed as:
\begin{equation}
\sigma_{ij} = \frac{\mu}{Re_{\infty}} \left( \frac{\partial u_i}{\partial x_j} +\frac{\partial u_j}{\partial x_i} -\frac{2}{3} \frac{\partial u_k}{\partial x_k}\delta_{ij} \right) \, ,
\label{eq:viscous_stress_tensor}
\end{equation}

\begin{equation}
E = \rho e + \frac{1}{2}\rho u_i^2 \, ,
\end{equation}

\begin{equation}
q_i = - \frac{1}{Re_{\infty} Pr_{\infty}} k_{th} \frac{\partial T}{\partial x_i} \, .
\end{equation}

Freestream conditions are indicated with subscript $(\cdot)_{\infty}$, and these are used for the non-dimensionalization of conservative and primitive variables \citep{Marxen2010}. The dimensional variables are marked with the symbol $\tilde{(\cdot)}$, whereas the latter is omitted for the non-dimensional form. Sutherland's law, with Sutherland's temperature $\tilde{T}_s=110.4K$ \citep{Anderson1989}, is used to compute molecular viscosity ($\mu$). Based upon the non-dimensionalization of the Navier-Stokes equations used, the Reynolds number ($Re_{\infty}$) and Prandtl number ($Pr_{\infty}$) formulation is

\begin{equation}
Re_{\infty} = \tilde{\rho}_{\infty} \tilde{c}_{\infty} \tilde{L}_{ref} / \tilde{\mu}_{\infty} \, ,
\end{equation} 

\begin{equation}
Pr_{\infty} = \tilde{\mu}_{\infty} \tilde{c}_{p} / \tilde{k}_{th,\infty} \, ,
\end{equation} 
where $\tilde{\rho}_{\infty}$, $\tilde{c}_{\infty}$, $\tilde{\mu}_{\infty}$ and $\tilde{k}_{th,\infty}$ are the freestream density, speed of sound, dynamic viscosity and thermal conductivity, respectively, $\tilde{L}_{ref}$ ($=6.20\times 10^{-4}
  \mathrm{m}$) is the reference length scale, and $\tilde{c}_{p}$ is the specific heat at constant pressure. The three-dimensional velocity vector is indicated as $\left[u_1 \: u_2 \: u_3 \right]^T = \left[u \: v \: w \right]^T$, and it is a function of the spatial coordinates $\left[x_1 \: x_2 \: x_3 \right]^T = \left[x \: y \: z \right]^T$. Here, $x$, $y$, and $z$ are non-dimensional spatial coordinates defined as
$x=\tilde{x}/\tilde{L}_{ref}$, $y=\tilde{y}/\tilde{L}_{ref}$, and $z=\tilde{z}/\tilde{L}_{ref}$. The freestream Mach number is defined as $M_{\infty}=\tilde{u}_{\infty}/{\tilde{c}_{\infty}}$.

In the rest of the manuscript, velocity and temperature scales are normalized with the freestream velocity, $\tilde{u}_{\infty}$, and static temperature, $\tilde{T}_{\infty}$, respectively. In place of the non-dimensional streamwise coordinate, $x$, a local Reynolds number, $Re_x=\sqrt{xRe_{\infty}M_{\infty}}$, is sometimes also used. The ratio of the specific heats ($\gamma$) is set to $\gamma=1.4$, and $Pr_{\infty}=0.71$. Quantities evaluated at the wall are indicated in the rest of the manuscript with the subscript $w$.

The structure and methods used for the DNS solver closely follow the algorithm described by \citet{Nagarajan2003} and \citet{Nagarajan2007}. The equations are discretized on a spatially structured, curvilinear grid with a staggered approach for the conservative variables. As further detailed in \citet{Marxen2010}, for the spatial discretization and the time integration a nominally $6^{th}$ order compact finite difference scheme and an explicit $3^{rd}$ order Runge-Kutta method are used, respectively. During the transition process, the dynamics of compressible turbulence is highly non-linear and can contribute to numerical instability via aliasing errors \citep{Song2024}. To preserve numerical stability of the high-order spatial discretization scheme, the  approximate deconvolution model proposed by \citet{Stolz2001} is used when assembling the numerical fluxes. The compressible DNS solver has been extensively used and verified for the computation of linear (small-amplitude) and non-linear evolution of boundary layer disturbances \citep{Marxen2011}, with \citep{Marxen2013} and without \citep{Marxen2010} high-temperature gas effects.

\subsubsection{Computational settings}\label{sec:domain}
\begin{figure}
\centering
\includegraphics[width=\textwidth]{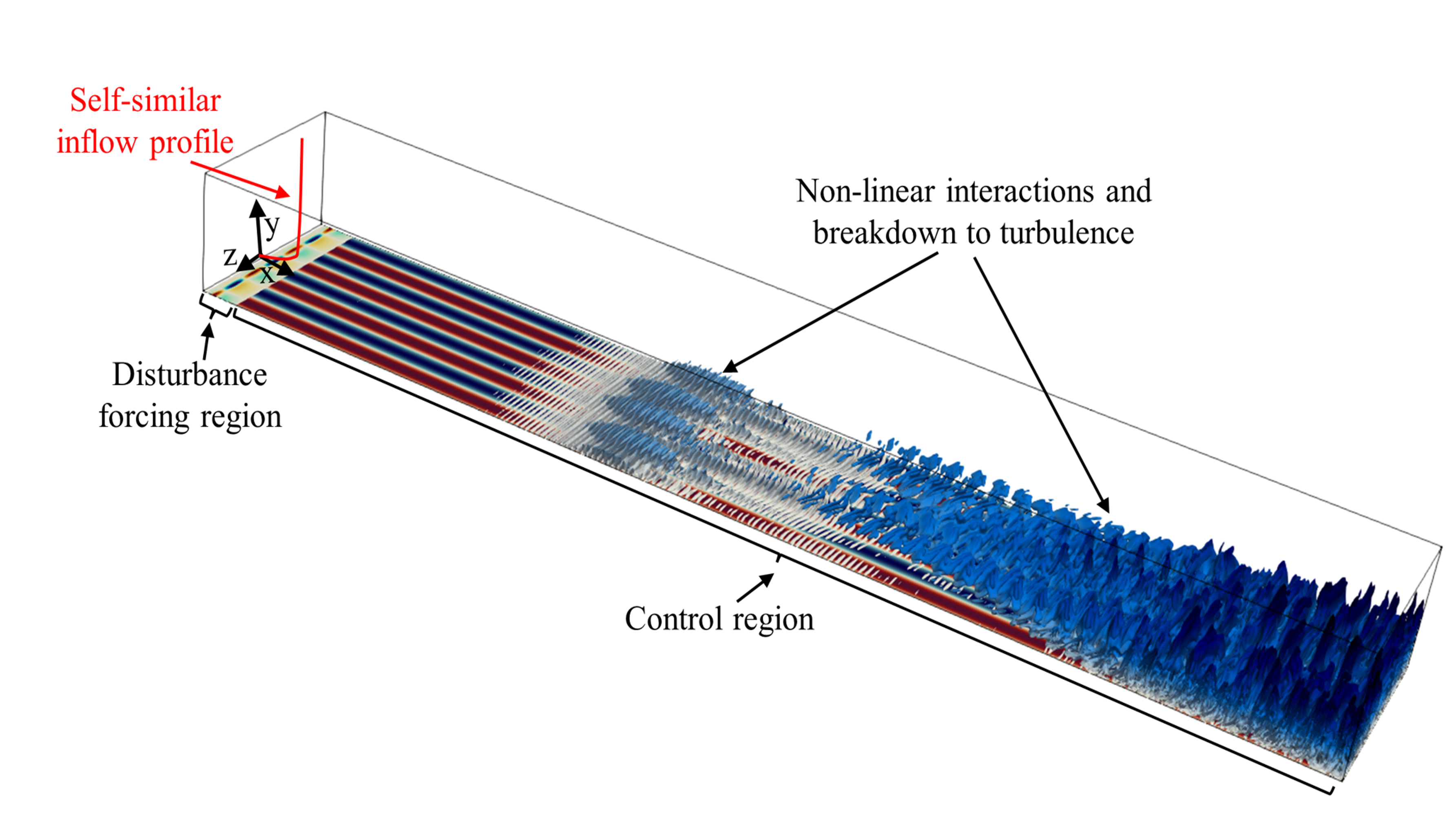}
\caption{Schematic diagram of computational domain for DNS computations for hypersonic boundary layer transition control under deterministic forcing. The control region is applied as a boundary condition at the wall and is indicated by the alternate red-blue spanwise pattern. Deterministic forcing is applied as a blowing-suction boundary condition and indicated by the label disturbance forcing region. Transition to turbulence is depicted with white and blue isosurfaces of Q-criterion.}
\label{fig:dns_domain}
\end{figure}

A schematic diagram of the computational domain for the DNS computations that capture transition to turbulence is depicted in Fig. \ref{fig:dns_domain}. A self-similar, laminar solution is injected at domain inlet and it develops along a viscous, isothermal wall. Sponge regions are used at the inlet, outlet and upper boundary to damp the solution towards a self-similar laminar state and prevent spurious reflection of pressure waves inside the domain. Periodic boundary conditions are applied in the spanwise direction at both sides of the domain. Transition to turbulence is promoted through a disturbance forcing region at the wall and further details are given in section \ref{sec:actuator}.

The streamwise, $x$, wall-normal, $y$, and spanwise, $z$, extent of the domain are indicated as $L_x$, $L_y$ and $L_z$, respectively, and are selected based on previous studies \citep{Franko2013,Franko2014}. The domain starts at $x=x_s$ and terminates at $x=x_e$, and is discretized through $N_x$ and $N_z$ uniformly spaced grid nodes in the streamwise and spanwise directions, respectively. In the wall-normal direction, the following stretching formula is used

\begin{equation}
y(m) = \frac{4}{5}\left(\frac{m-1}{N_y-1}\right)^3 + \frac{1}{5} \frac{m-1}{N_y-1} \, ,
\end{equation}
where $m$ is the grid node index and $N_y$ is the total number of grid points in the wall normal direction. The extent of the computational domain is determined based on previous studies \citep{Franko2013,Franko2014}. The number of grid points within the sponge regions is indicated with $N_{sponge}$, and further details are provided in section \ref{sec:results}. 

Two laminar to turbulence transition scenarios are investigated, namely second Mack mode fundamental resonance (SMF) and first Mack mode oblique (FMO) breakdown \citep{Franko2013}. The choice of the timestep for both configurations is based upon previous work \citep{Franko2013,Franko2014}. A fundamental period, $\tau_0=2\pi/\omega_0$, is used for non-dimensionalization purposes and is based upon the fundamental angular frequency ($\omega_0$) of the forcing disturbance. The details about operating conditions, domain and grid size, and number of timesteps ($N_{t}$) per fundamental period are introduced in section \ref{sec:results}.

\subsubsection{Disturbance forcing}\label{sec:actuator}
In this section, the wall boundary condition used to force transition to turbulence is described, and the mathematical definition, amplitude and frequency characteristics are largely based on \citet{Franko2013}.
Two breakdown to turbulence scenarios via deterministic forcing are investigated and referred to as second Mack mode fundamental resonance (SMF) and first Mack mode oblique (FMO) breakdown \citep{Franko2013}. Wall blowing and suction is used to force transition to turbulence within the computational domain; this boundary condition is sometimes referred to as actuator in the rest of the manuscript. The wall-normal profile implemented as the boundary condition in the disturbance forcing region is expressed as
\begin{equation}
\begin{cases}
    \frac{\tilde{v}_{w}}{\tilde{c}_{\infty}} = {v}_{w} = \exp\left(-\frac{\sqrt{2}}{2} 	\xi^2\right) g(z) \sum_i A_i \sin \left( \omega_i t - k_i z \right)\\
    g(z) = \left[1.0 + 0.1\left(\exp \left[-\left(\frac{z-2\pi}{2\pi}\right)^2\right] - \exp \left[-\left(\frac{z+2\pi}{2\pi}\right)^2\right]\right)\right]\\
   \xi = \frac{x - x_{c,strip}}{L_{strip}}  
\end{cases}  \, ,
\label{eq:actuator_eq_FL}
\end{equation}
where $A_i$, $\omega_i$ and $k_i$ are the non-dimensional amplitudes, angular frequencies and spanwise wavenumbers of the wall normal disturbances, with values based upon \citet{Franko2013} studies and summarized in table \ref{tab:transition_forcing}. The subscript $(\cdot)_w$ denotes quantities evaluated at the wall. The function $g(z)$ in equation \eqref{eq:actuator_eq_FL} introduces spanwise asymmetry and promotes faster transition to turbulence \citep{Wang1996}. The center of the actuator is at $x_{c,strip}=17.5$ and its length $L_{strip}=5$. 

\begin{table}
\begin{ruledtabular}
  \caption{Summary of the parameters for the disturbance forcing law for the two breakdown to turbulence mechanisms investigated, second Mack mode fundamental frequency (SMF) and first Mack mode oblique breakdown (FMO).}
  \label{tab:transition_forcing}
  \begin{tabular}{ccccc}
      Case & Mode (i) & $A_i [\times     M_{\infty}]$ & $\omega_i [\times 2\pi/\omega_0]$ & $k_i [\times L_z/2\pi]$ \\[3pt]
      SMF & 1 & 0.1   & 1   & 0         \\
      SMF & 2 & 0.005 & 1/2 & 0         \\
      SMF & 3,4 & 0.005 & 0   & {[}1,2{]} \\
      SMF & 5,6 & 0.005 & 1/2   & $\pm1$    \\
      SMF & 7,8 & 0.005 & 1   & $\pm2$    \\
      SMF & 9,10 & 0.005 & 1   & $\pm1$    \\ \hline
      FMO & 1,2 & 0.05  & 1   & $\pm3$    \\ 
  \end{tabular}

  \end{ruledtabular}
\end{table}

\subsubsection{Wall temperature boundary condition}\label{sec:tw}
This section provides a description of the wall temperature boundary condition that is used to generate the heated streaks to stabilize the boundary layer and delay transition to turbulence. The mathematical definition is largely based on our previous work \citep{Boscagli2025_SCITECH, Boscagli2026_JFM}. The wall temperature boundary condition is  
\begin{equation}
    T_w = T_{w,base} \left(1 + A_{T_{w}} \sin \left( k_s z + \theta \right) \right) \, ,
\label{eq:Tw_eq}
\end{equation}
where $A_{T_w}$ sets the amplitude of the wall temperature variation  relative to the baseline (uniform) wall temperature, $k_s=2\pi/\lambda_{z,s}$ sets the fundamental spanwise wavenumber of the control streaks, and $\theta \in [0,2\pi]$ is used to control the relative phase between the wall temperature and the disturbance forcing boundary conditions. The parameter $\theta$ is set to zero for most of the investigations, but for the studies presented in appendix \ref{app:phase_effect_SMF}, that investigate the effect of changes in spanwise phase on the control method effectiveness.

\subsubsection{Time-advancement and data analysis methods}\label{sec:converegence}
For the time-advancement, 1200 timesteps per second Mack mode fundamental period ($\tau_0 = 2\pi/\omega_0 \approx 2.4$) are used. To achieve statistical convergence, the computations for the uncontrolled configurations are advanced in time for approximately 1.5 convective times ($\tilde{L}_x/\tilde{u}_{\infty}$). For the controlled configurations, the computations are initialized from the uncontrolled case and using a linear temporal ramp-up of $A_{T_w}$ for the first 15-30$\tau_0$. This is done in steps of $\Delta A_{T_w}=0.05$, and the simulations are run for approximately 1.5 convective times. Upon statistical convergence of both the uncontrolled and controlled configurations, data are sampled for $5\tau_0$ with a sample frequency $f_{Ny}=12/\tau_0$. This, along with the spanwise resolution of the grid, provides sufficient spectral resolution and range for the frequency ($f_n$), spanwise wavenumber ($k_m$) analysis. The averaging-window length also guarantees sufficiently converged turbulence statistics (further details are in appendix~\ref{app:convergence}). In the following, $f_n$ and $k_m$ are normalized as $f = f_n\tau_0$ and $k = k_m \frac{L_z}{2\pi}$.
The computational model employs periodic boundary conditions in the spanwise direction, and therefore a frequency ($f$) and spanwise wavenumber ($k$) Fourier decomposition of the primitive variables is used. The Chu's energy ($E_{Chu}^{fk}$, \citep{Chu1965}) is used to track the evolution of the boundary layer instabilities and it is defined as 

\begin{equation}
\begin{split}
E_{Chu}^{fk}(x) = \frac{1}{2} \int_{0}^{L_y} \Biggl[ \overline{\rho}\left( \hat{u}\hat{u}^* + \hat{v}\hat{v}^* + \hat{w}\hat{w}^* \right)
                  + \frac{\overline{T}}{\gamma M_{\infty}^2 \overline{\rho}} \hat{\rho}\hat{\rho}^* 
                  + \frac{\overline{\rho}}{\gamma \left( \gamma - 1 \right) M_{\infty}^2 \overline{T}}\hat{T}\hat{T}^* \Biggr] dy \, .          
\end{split}
\label{eq:Echu}
\end{equation}
In equation \eqref{eq:Echu}, $\overline{(\cdot)}$, $(\cdot)'$ and $(\hat{\cdot})$ indicate the mean flow deformation, the amplitude of the fluctuations and the Fourier coefficient, respectively, and $(\cdot)^*$ indicates the complex conjugate. $L_y$ indicates the wall-normal extent of the computational domain. The Chu's energy is chosen as a metric to quantify the modal energy as this takes into account both kinetic and thermodynamic energy contributions \citep{Unnikrishnan2020,Guo2023}, which are both relevant in the present study where streaks are generated through manipulation of the surface temperature. In addition, the Chu's energy it is also a commonly used metric in compressible linear input/output analysis \citep{Bugeat2019}, for the study of modal and non-modal boundary layer linear stability. The amplitude of the control streaks with $k_s$ wavenumber is quantified based on $|\hat{u}|^{0k_s}_{max}$, in place of $As_u$ introduced by \citet{Andersson2001}. Compared to $As_u$, the definition based on the frequency-wavenumber decomposition of the streamwise velocity field decouples the amplitude of the control-streaks from the amplitude of streaks of different wavelength that may be part of the forcing disturbance or the result of triadic interaction of oblique waves. In experiments, $As_u$ is generally used due to the limited temporal and spatial resolution, which is needed for the $(f,k)$ decomposition. Thus, in the rest of the text, provided that there is no ambiguity in the definition of the control streaks, for conciseness the two notations are used interchangeably. 

Boundary layer transition from laminar to turbulent is determined based on the local skin friction coefficient, defined as
\begin{equation}
C_f=\frac{2}{Re_{\infty}M_{\infty}} \left(\mu\frac{\partial u}{\partial y}\right)_w=\frac{2\tau_w}{Re_{\infty}M_{\infty}} \, .
\end{equation}

For the analysis of the late stage of transition to turbulence, Favre-averaging \citep{Wilcox1998} is used for the computation of velocity and temperature fluctuations. Favre-averaged mean and fluctuating quantities are indicated with $(\breve{\cdot})$ and $(\cdot)^{''}$, respectively, while time-averaging is indicated with $\langle \cdot \rangle$. The effect of control streaks on the late stage of transition to turbulence is further investigated through the turbulent kinetic energy, defined as $\kappa \equiv \frac{1}{2}\,\breve{u_i^{''}u_i^{''}}$, or equivalently $\langle \rho \rangle \kappa = \frac{1}{2}\,\langle \rho\,u_i^{''}u_i^{''}\rangle$, and through the turbulent heat flux $\langle \rho u_i^{''} T^{''}\rangle$. The budget equation for the Favre-averaged turbulent kinetic energy is written as

\begin{equation}
\frac{\partial (\langle \rho \rangle \kappa)}{\partial t}
+
\langle \rho \rangle \breve{u}_j \frac{\partial \kappa}{\partial x_j}
= \mathcal{P} - \mathcal{D} + \mathcal{T}_{\mu} + \mathcal{T}_t + \mathcal{T}_p + \Pi_w + \Pi_d \, ,
\label{eq:FA_TKE_budget}
\end{equation}
where the terms on the right-hand side represent production ($\mathcal{P}$), destruction ($\mathcal{D}$), molecular diffusion ($\mathcal{T}_{\mu}$), turbulent transport ($\mathcal{T}_t$), pressure transport ($\mathcal{T}_p$), pressure work ($\Pi_w$), and pressure dilatation ($\Pi_d$), respectively. Following \citet{Wilcox1998}, these terms are defined as

\begin{equation}
\mathcal{P} = - \langle \rho u_i^{''} u_j^{''} \rangle \frac{\partial \breve{u}_i}{\partial x_j} \, ,
\end{equation}

\begin{equation}
\mathcal{}{D} = \left\langle \sigma_{ij}^{''} \frac{\partial u_i^{''}}{\partial x_j} \right\rangle \, ,
\label{eq:TKE_D}
\end{equation}

\begin{equation}
\mathcal{T}_t = - \frac{\partial }{\partial x_j} \left( \left\langle \rho u_j^{''} \frac{1}{2} u_i^{''}u_i^{''} \right\rangle \right), \qquad
\mathcal{T}_p = - \frac{\partial }{\partial x_j} \left( \left\langle p^{'} u_j^{''} \right\rangle \right), \qquad
\mathcal{T}_{\mu} = \frac{\partial }{\partial x_j} \left( \left\langle \sigma_{ji}^{''} u_i^{''} \right\rangle \right) \, ,
\end{equation}

\begin{equation}
\Pi_w = - \langle u_i^{''} \rangle \frac{\partial \langle p \rangle}{\partial x_i}, \qquad
\Pi_d = \left\langle p^{'} \frac{\partial u_i^{''}}{\partial x_i} \right\rangle \, .
\end{equation}
Here, $-\langle \rho u_i^{''} u_j^{''} \rangle$ is the Favre-averaged Reynolds stress tensor, and $\sigma_{ij}^{''}$ is the fluctuating viscous stress tensor defined by equation \eqref{eq:viscous_stress_tensor}. For convenience, the pressure-related terms are also considered in combined form as
\begin{equation}
\Pi \equiv \Pi_w + \Pi_d .
\end{equation}

For the characterization of the smallest dissipative scales, the Kolmogorov length scale is then defined as
\begin{equation}
\eta = \left( \frac{\nu^3}{\varepsilon} \right)^{1/4} \, .
\label{eq:kolmogorov_length scale}
\end{equation}
In the preceding equation, $\nu = \mu/\rho$ is the local kinematic viscosity, and $\varepsilon$ is the turbulent dissipation rate, which is defined as 
\begin{equation}
\varepsilon = 2\nu\,S_{ij}^{''}S_{ij}^{''},
\end{equation}
with fluctuating strain-rate tensor
\begin{equation}
S_{ij}^{''}
=
\frac{1}{2}
\left(
\frac{\partial u_i^{''}}{\partial x_j}
+
\frac{\partial u_j^{''}}{\partial x_i}
\right).
\end{equation}
In the present analysis, $\nu$ and $\varepsilon$ are evaluated from the instantaneous flow field and subsequently averaged in time. To assess the degree of rarefaction at the smallest dynamically relevant scales, a Knudsen number based on the Kolmogorov length scale is introduced as
\begin{equation}
Kn_{\eta} = \frac{\lambda}{\eta} \, ,
\label{eq:knudsen}
\end{equation}
where $\lambda$ is the local mean free path. For a calorically perfect gas, $\lambda$ is estimated as
\begin{equation}
\lambda = \frac{\mu}{p}\sqrt{\frac{\pi R T}{2}} \, .
\end{equation}
For the present cases, $Kn_{\eta}$ remains below unity ($\,1\times10^{-8}\lesssim Kn_{\eta}\lesssim2.9\times10^{-1}$). While localized regions closer to the wall reach the slip/early-transition range ($Kn_{\eta}\gtrsim 0.1$), the bulk of the flow remains in the continuum regime, and previous studies at comparable global conditions (global Knudsen number $Kn= \sqrt{\pi\gamma/2}/Re_{\infty}\sim10^{-3}$) suggest limited impact on the turbulence cascade statistics \citep{Gallis2021}. Further details are provided in Appendix~\ref{app:verification}.

\section{Results}\label{sec:results}
DNS studies of laminar to turbulence transition under deterministic forcing (section \ref{sec:actuator}) are used to assess the effectiveness of spanwise non-uniform surface temperature to delay transition. Laminar to turbulence transition via second Mack mode fundamental resonance (SMF) and first Mack mode oblique breakdown (FMO) is investigated for a Mach 6, cold ($T_{w,base}=5$) flat plate boundary layer with zero pressure gradient. Further details about the operating conditions, domain size and grid resolution are summarized in tables \ref{tab:transition_op_conditions} and \ref{tab:transition_grid}. In table~\ref{tab:transition_grid}, the domain size is expressed both relative to the reference length scale ($\tilde{L}_{ref}$) as well as relative to non-dimensional boundary layer thickness at the inlet of the domain ($\delta_{99,in}$). The Reynolds-number range was selected to be consistent with prior high-fidelity studies of deterministic transition and modal breakdown \citep{Franko2013}. The corresponding unit Reynolds number ($Re_{unit}=\tilde{\rho}_{\infty}\tilde{u}_{\infty}/\tilde{\mu}_{\infty}$) lies within the operational range of high-density hypersonic facilities \citep{Schneider2008}; experimental evidence indicates that unit Reynolds number is strongly correlated with transition behavior in hypersonic boundary layers \citep{Anderson1989}. Furthermore, previous analyses at similar Mach number \citep{Franko2014} show that adverse-pressure-gradient effects amplify linear growth and advance transition onset, while the dominant nonlinear deterministic breakdown routes remain qualitatively similar to the zero-pressure-gradient case. Thus, the present study focuses on a qualitative and quantitative assessment of transition-control physics with thermally generated streaks under canonical, zero-pressure-gradient conditions, while the coupled influence of adverse-pressure-gradient effects and control-streak forcing is an important extension left for future work.

\begin{table}
  \begin{ruledtabular}
  \caption{Summary of operating and boundary conditions for the transition to turbulence studies.}
  \label{tab:transition_op_conditions}      
\def~{\hphantom{0}}
  \begin{tabular}{cccccccccc}
     Case & $M_{\infty}$ & $\tilde{T}_{\infty} [\textnormal{K}]$ & $T_{w,base}$ & $(Re_{\infty}M_{\infty})$ & $Re_{unit} [\textnormal{1/m}]$ & $Re_{x_{s}}$ & $Re_{x_{e}}$  & $Re_{x_{c,strip}}$ & $L_{strip}$ \\ [3pt]
     SMF & $6.0$ & $65.15$ & $5$ & $4333$ & $7.2\times 10^{6}$ & $0$  & $2550$  & $275$ & $5$ \\
     FMO & $6.0$ & $65.15$ & $5$ & $4333$ & $7.2\times 10^{6}$ & $0$  & $1862$  & $275$ & $5$ \\
  \end{tabular} 
  \end{ruledtabular}
\end{table}

\begin{table}
  \begin{ruledtabular}
  \caption{Summary of computational grid details.}
  \label{tab:transition_grid}
  \begin{tabular}{ccccccccc}
     Case &
     \begin{tabular}{@{}c@{}}$[x_s,x_e]$\\($L_x/\delta_{99,\mathrm{in}}$)\end{tabular} &
     \begin{tabular}{@{}c@{}}$L_y$\\($L_y/\delta_{99,\mathrm{in}}$)\end{tabular} &
     \begin{tabular}{@{}c@{}}$L_z$\\( $L_z/\delta_{99,\mathrm{in}}$)\end{tabular} &
     $N_x \times N_y \times N_z$ &
     $[N_x,N_y]_{\mathrm{sponge}}$ &
     $\Delta x_{\max}^{+}$ &
     $\Delta y_{w,\max}^{+}$ &
     $\Delta z_{\max}^{+}$\\[3pt]

     SMF &
     \begin{tabular}{@{}c@{}}$[0,1500]$\\$(\sim967)$\end{tabular} &
     \begin{tabular}{@{}c@{}}$50$\\$(\sim18)$\end{tabular} &
     \begin{tabular}{@{}c@{}}$50.27$\\$(\sim33)$\end{tabular} &
     $2560 \times 150 \times 128$ & $[24,32]$ & $6.78$ & $0.78$ & $4.54$\\

     FMO &
     \begin{tabular}{@{}c@{}}$[0,800]$\\$(\sim543)$\end{tabular} &
     \begin{tabular}{@{}c@{}}$50$\\$(\sim19)$\end{tabular} &
     \begin{tabular}{@{}c@{}}$62.83$\\$(\sim44)$\end{tabular} &
     $2048 \times 150 \times 256$ & $[24,32]$ & $3.24$ & $0.89$ & $5.16$\\
  \end{tabular}
  \end{ruledtabular}
\end{table}

The computational methodology was previously verified \citep{Boscagli2025_SCITECH} with the existing data in the literature \citep{Franko2013} for a nearly adiabatic flat plate ($T_{w,base}=6.5$) for $M_{\infty}=6$. Typical DNS grid requirements \citep{Franko2013} are met. In the laminar region there are approximately 24 (streamwise) points per second Mack mode fundamental wavelength for the SMF configuration, and approximately 85 points per spanwise wavelength of the first Mack mode for the FMO case. In the turbulent region, the grid resolution is evaluated based on friction length $\delta_{\nu} = \nu_w/u_{\tau}$ (with $u_{\tau}=\sqrt{\tau_w/\rho_w}$), and expressed in near wall units indicated with subscript $(\cdot)^+$. The maximum grid spacing in the streamwise ($\Delta x^{+}_{max}$), spanwise ($\Delta z^{+}_{max}$) and wall-normal ($\Delta y^{+}_{max}$) directions for both the SMF and FMO configurations are reported in table \ref{tab:transition_grid}, and typical DNS grid requirements based on the existing literature \citep{Franko2013} are met. Sections \ref{sec:transition_SMF} and \ref{sec:transition_FMO} below present the assessment of the  uncontrolled (baseline) and controlled configurations for both the SMF and FMO configurations, respectively.

\subsection{Second Mack mode fundamental resonance}\label{sec:transition_SMF}
\begin{figure}
\centering
\includegraphics[trim={0cm 3cm 0cm 3cm},clip,width=\textwidth]{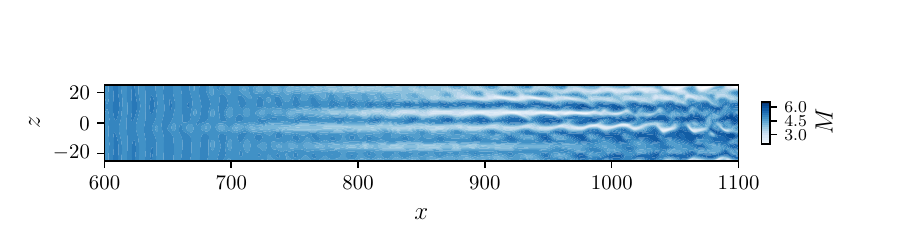}
\caption{DNS analysis of the baseline, uncontrolled SMF case; instantaneous distribution of streamwise Mach number at $y=4.5$. For better visualization, the figure aspect ratio is distorted by $20\%$, and only $x\in[600,1100]$ is shown.}
\label{fig:uncontrolled_mach_SMF}
\end{figure}

Breakdown to turbulence via second Mack mode fundamental resonance is investigated for a Mach 6 boundary layer over a flat plate configuration. The axial Mach number distribution across the streamwise and spanwise directions at $y=4.5$ in Fig. \ref{fig:uncontrolled_mach_SMF} provides an indication of the flow topology for the uncontrolled (baseline) configuration. Downstream of the energy peak of the second Mack mode ($x \approx 650$), there is formation of large amplitude steady streaks with the most energetic harmonic being the $(f,k)=(0,4)$ based on a frequency-wavenumber decomposition. This is not part of the spectrum of the forcing disturbance law (table \ref{tab:transition_forcing}), but is the result of triadic, difference interaction of the disturbance oblique waves (mode 4 in table \ref{tab:transition_forcing}). 

\begin{figure}
\centering
\includegraphics[trim={0cm 0cm 0cm 0.0cm},clip,width=\textwidth]{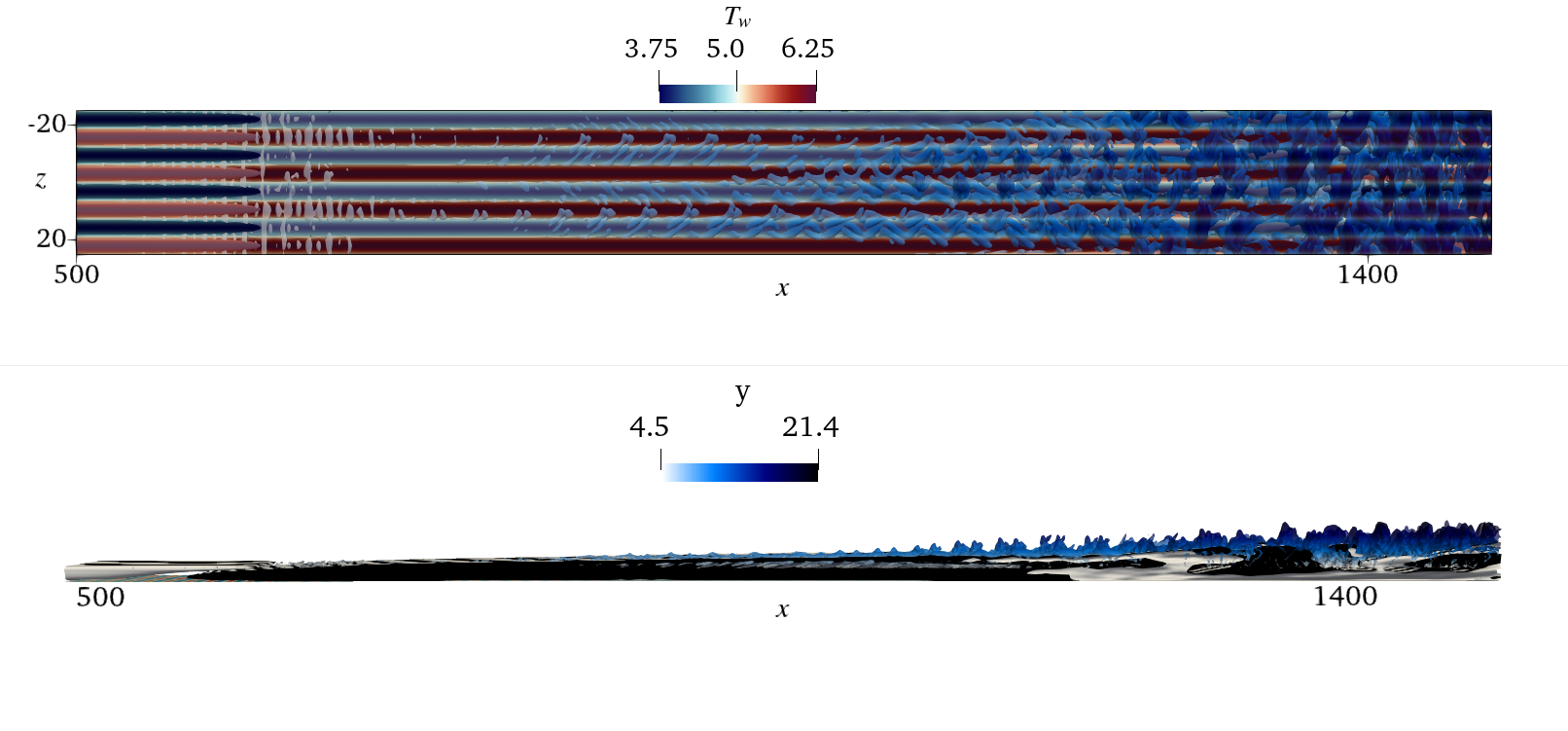}
\caption{DNS analysis of the controlled SMF case. Top (upper) and front (bottom) view of  transition to turbulence via second Mack mode fundamental resonance. $M_{\infty}=6$, $Re_{unit}=7.2 \times 10^6$1/m, $T_{w,base}=5$, $A_{T_w}=0.25$, $k_s=4$. Isosurfaces of Q-criterion ($10^{-8}$) colored with wall normal distance ($y)$. Black and white isosurfaces ($u^{04}=\pm 0.01$) depict instead high and low speed streaks, respectively. For better visualization, the figure aspect ratio is set to 2.}
\label{fig:SMF_controlled}
\end{figure}

For the controlled configuration, a spanwise non-uniform surface temperature with $A_{T_w}=0.25$ and $k_{s}=2\pi/\lambda_{z,s}=4$ is enforced in the computational domain to understand whether transition delay can be achieved via either constructive or destructive interference of the large amplitude streaks due to non-linear interactions and the control-streaks (Fig. \ref{fig:SMF_controlled}). The spanwise variation in surface temperature corresponds to approximately $162$ K. The choice of the fundamental wavenumber of the control-streaks ($k_s$) is based upon previous studies  \citep{Boscagli2025_SCITECH}. This corresponds to a wavelength that is approximately $9$ times the boundary layer thickness at the inlet of the computational domain ($\delta_{99,in}$). For these operating conditions, it is expected to provide a near-optimum stabilization of the second Mack mode, via MFD in the initial linear stage of transition. The choice of the wavelength is therefore made independently of the wavelength of the streaks that are generated via non-linear interaction subsequent to the initial planar wave amplification.

\begin{figure}
  \centering 
  \subfloat[]{\includegraphics[width=0.48\textwidth]{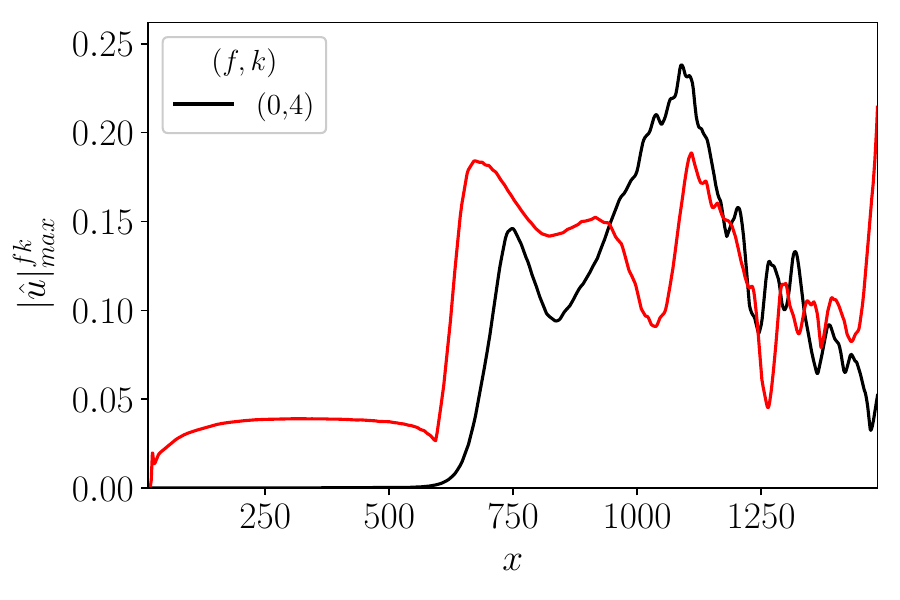}\label{fig:SMF_controlled_Fourier_streaks}}
\hfill
  \subfloat[]{\includegraphics[width=0.48\textwidth]{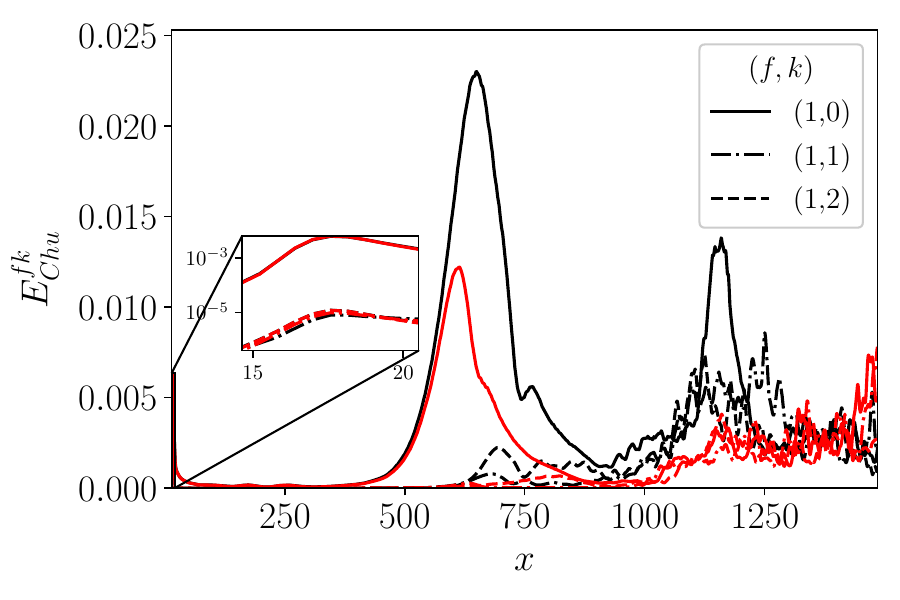}\label{fig:SMF_controlled_Fourier_second_mode}}
\caption{DNS analysis of the SMF case; streamwise distribution of the amplitude of (\textit{a}) streaks harmonics and (\textit{b}) second Mack mode energy. Black lines: uncontrolled configuration; red lines: controlled configuration, $A_{T_w}=0.25$. The inset in (\textit{b}) depicts the modal energy introduced by the blowing and suction at the second Mack mode fundamental frequency, and it is the same for the uncontrolled and controlled configuration.}
\label{fig:SMF_controlled_Fourier}
\end{figure}

The spanwise non-uniform surface temperature generates control-streaks with a maximum amplitude of approximately 3.5\% of the freestream speed ($u_{\infty}$). This is quantified through the amplitude of the $(f,k)=(0,4)$ Fourier harmonic of the streamwise velocity in Fig. \ref{fig:SMF_controlled_Fourier_streaks} for $x<500$. Further downstream ($x>500$), the amplitude of the $(f,k)=(0,4)$ harmonic rapidly grows, and it peaks at around $x\approx650$ to approximately 20\% (fig. \ref{fig:SMF_controlled_Fourier_streaks}). The comparison with the uncontrolled configuration indicates the presence of constructive interference between the control-streaks and the streaks due to non-linear interactions. The control-streaks stabilize the second Mack mode with a significant reduction in the amplitude of the modal energy of the fundamental, 2D, harmonic, depicted by the $(f,k)=(1,0)$ harmonic in Fig. \ref{fig:SMF_controlled_Fourier_second_mode} for $x\approx 600$. This is in agreement with our previous studies \citep{Boscagli2026_JFM}.

\begin{figure}
  \centering 
  \includegraphics[width=\textwidth]{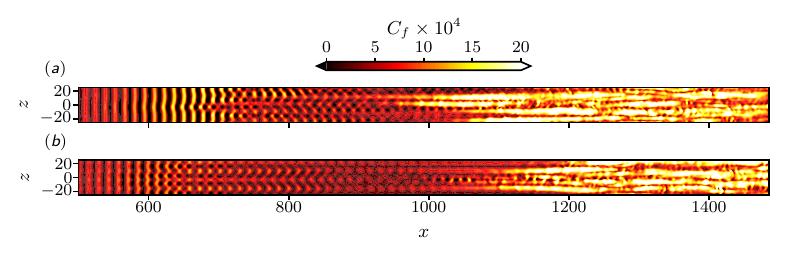}
\caption{DNS analysis of the SMF case; instantaneous snapshots of skin friction coefficient for (\textit{a}) uncontrolled and (\textit{b}) controlled ($A_{T_w}=0.25$, $k_s=4$) configuration.}
\label{fig:SMF_controlled_Cf_contour}
\end{figure}

Fig. \ref{fig:SMF_controlled_Cf_contour} shows the instantaneous distribution of skin friction coefficient for both the uncontrolled and controlled configurations. At around $x \approx 700$, the control method significantly reduces the peak value of $C_f$ associated to the high-frequency fluctuations of the second Mack mode. Further downstream, for the controlled configuration, the large amplitude streaks due to non-linear interactions and constructive interference with the control-streaks stay coherent for a longer streamwise distance (Fig. \ref{fig:SMF_controlled_Cf_contour}). In addition, the signature of the streaks on the instantaneous skin-friction coefficient indicates that the control method may contribute to preserve the left-right symmetry of the flow. This is further investigated in appendix \ref{app:symmetry} through the analysis of the streamwise evolution of the helicity ($h$), which is used as a measure of the structural coherence in the boundary layer. 

Previous studies that investigated the effect of the streaks generated through blowing and suction on first \citep{Sharma2019} and second Mack mode oblique breakdown \citep{Zhou2023} for supersonic and hypersonic configurations showed that the stabilization effect is mainly driven by the mean flow modification. Relative to the stabilization of the second Mack mode planar waves, \citet{Paredes2019} showed that the three-dimensional modulation of the boundary layer due to finite-amplitude ($As_u>10\%$) control streaks also plays significant role, and therefore three-dimensional effects need careful investigation. For weak control streaks ($As_u<5\%$), \citet{Boscagli2026_JFM} showed that the MFD drives the stabilization of the second Mack mode planar wave. In appendix \ref{app:phase_effect_SMF}, the assessment of the effect of the spanwise phase ($\theta$ in \eqref{eq:Tw_eq}) on the control method effectiveness is further investigated. It is shown that the control method is spanwise invariant to changes in $\theta$, and this is further indication that for this configuration the MFD dominates the stabilization mechanism.

\begin{figure}
\centering
\includegraphics[width=0.48\textwidth]{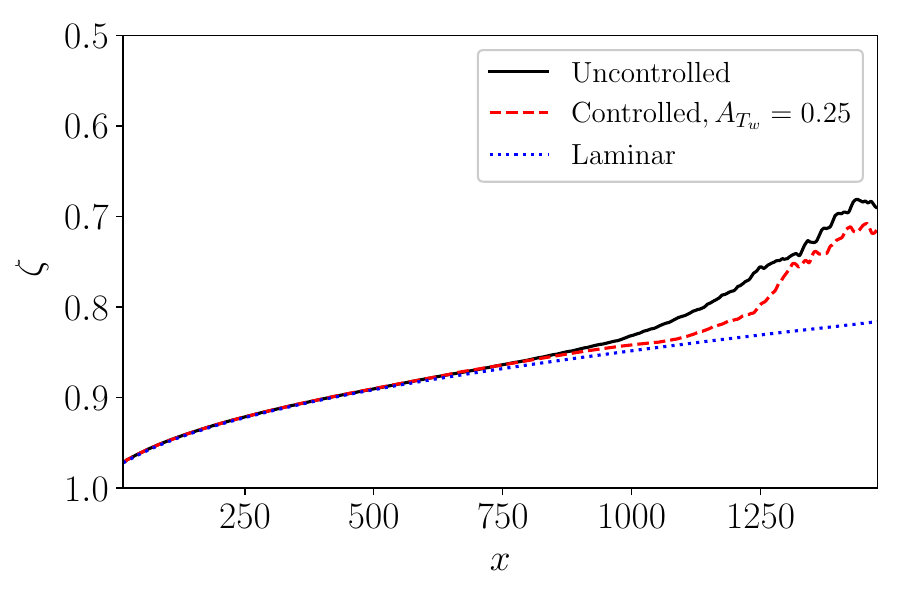}
\caption{Laminar and transitional DNS analysis for SMF case; effect of control method on streamwise distribution of the entropy function integrated on the cross-section.}
\label{fig:SMF_controlled_efficiency}
\end{figure}

The beneficial effects of the control method on the reduction of total pressure and temperature losses via a delayed transition to turbulence are further evaluated based on an isentropic efficiency, which is quantified with the entropy function ($\zeta$), and defined as

\begin{equation}
\zeta = \frac{1}{L_y L_z}\int \int \exp\left[ \frac{ -\left(\tilde{s} - \tilde{s}_{\infty} \right)}{\tilde{R}_{gas}} \right] dydz \, ,
\end{equation}
where $\tilde{R}_{gas}$ is the gas constant, $\tilde{s}$ is the local entropy, and $\tilde{s}_{\infty}$ is evaluated based on the freestream temperature and pressure. The control method delays the efficiency losses due to transition to turbulence further downstream compared to the uncontrolled configuration (Fig. \ref{fig:SMF_controlled_efficiency}). Comparison to a notionally optimal fully laminar configuration indicates that the benefit at the end of the computational domain remains marginal. 

\begin{figure}
  \centering 
  \subfloat[]{\includegraphics[width=0.48\textwidth]{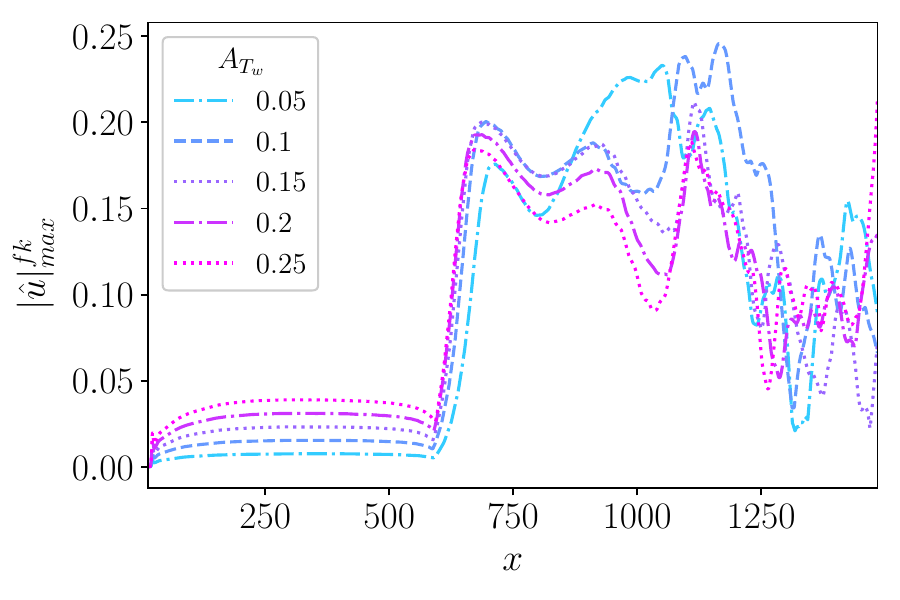}\label{fig:SMF_controlled_streaks}}  
\hfill
  \centering 
  \subfloat[]{\includegraphics[width=0.48\textwidth]{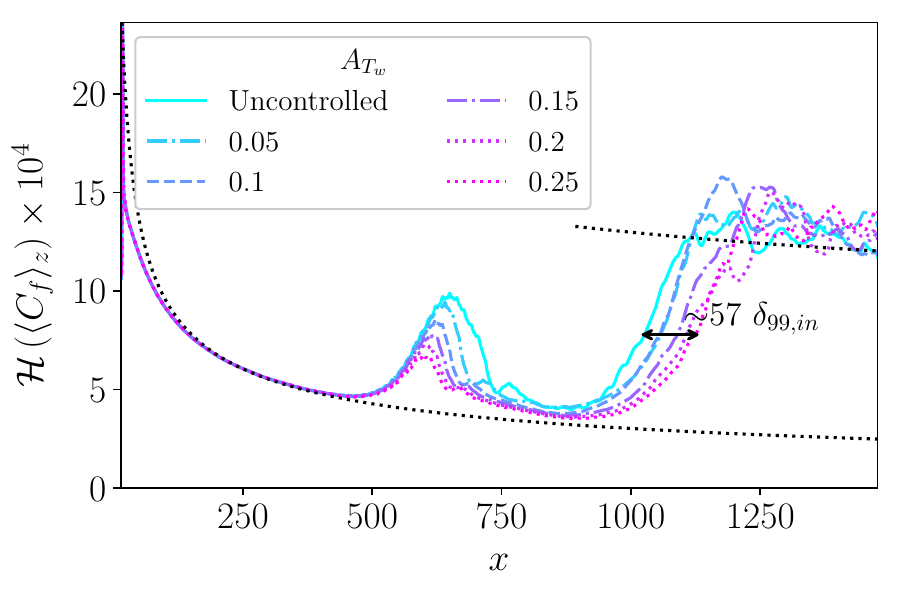}\label{fig:SMF_controlled_Cf_tzavg}}
\caption{DNS analysis of the SMF case; (\textit{a}) effect of $A_{T_w}$ on the amplitude of the control streaks ($x<500$). (\textit{b}) Effect of control method on the streamwise distribution of  the (time) envelope of the spanwise-averaged skin friction coefficient; grey dotted lines indicate the laminar and turbulent correlations.}
\label{fig:SMF_controlled_Atw_effect}
\end{figure}

To further understand the margin of improvement for the current method, the effect of a change in the amplitude of the spanwise temperature variation ($A_{T_w}$, equation \eqref{eq:Tw_eq}) is evaluated. A Hilbert transform ($\mathcal{H}(\cdot)$) of the spanwise-averaged skin friction coefficient ($\langle C_f \rangle_z$) is used to quantify the high-frequency (second Mack mode type) peak fluctuations associated with planar waves and evaluate the effect of an increase in streak amplitude via an increase in $A_{T_w}$ ($x<500$, Fig. \ref{fig:SMF_controlled_streaks}). For this case study, as $A_{T_w}$ is increased from $A_{T_w}=0.05$ to $0.25$, the spanwise temperature variation increase from approximately $32\textnormal{K}$ to $162\textnormal{K}$ and the amplitude of the streaks also increases from $\lvert \hat{u} \rvert_{max}^{04} \approx 0.75\%$ to approximately $3.8\%$ at the streamwise location of the onset of the second Mack mode planar wave, $(f,k)=(1,0)$ ($x\approx 500$ in Fig. \ref{fig:SMF_controlled_Fourier_second_mode}). At approximately the same streamwise location ($x\approx 500$) the analysis indicates that as the amplitude of the streaks increases, the peak shear-stress fluctuations and the streamwise extent of the second Mack mode amplification region reduce (Fig. \ref{fig:SMF_controlled_Cf_tzavg}). In addition, transition is also delayed further downstream. Relative to the uncontrolled case, for the controlled configuration with $A_{T_w}=0.25$ the amplitude and streamwise extent of the peak high-frequency shear stresses are reduced by approximately $30\%$ and $60\delta_{99,in}$, respectively. Based on the mid-point between the laminar and turbulent correlations, transition is delayed by approximately $57\delta_{99,in}$ for the controlled case with $A_{T_w}=0.25$ compared to the uncontrolled configuration.  

\begin{figure}
  \centering 
  \includegraphics[width=\textwidth]{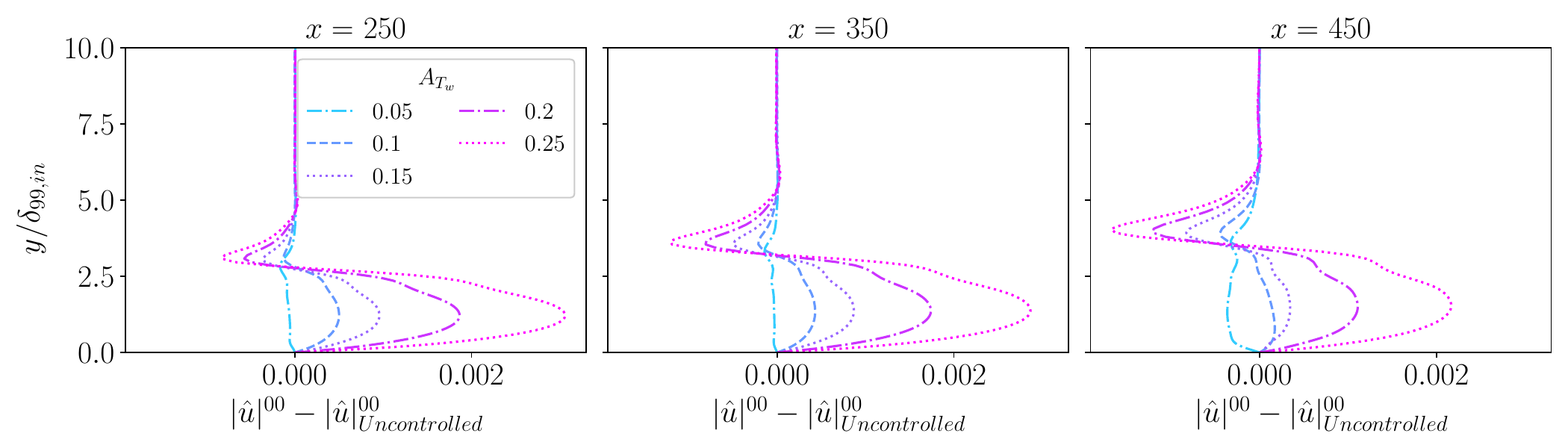}
\caption{DNS analysis of the SMF case; effect of $A_{T_w}$ on wall-normal distribution of the streamwise velocity mean flow deformation relative to the uncontrolled case at three axial locations ahead of the growth of the second Mack mode planar wave.}
\label{fig:SMF_controlled_Atw_effect_MFD}
\end{figure}

For similar characteristics (amplitude and wavelength) of the control streaks, \citet{Ren2016} concluded that the main stabilization mechanism of both first and second Mack modes is associated to a modification of the Mean Flow Deformation (MFD, $(f,k)=(0,0)$) that leads to a greater momentum boundary layer profile near the wall \citep{Cossu2002,Bagheri2007}. Previous work \citep{Boscagli2026_JFM} for a cold base flow modified by the thermally generated streaks, confirmed the driving role of the MFD on the stabilization of the second Mack mode using fully laminar DNS studies aided by linear stability theory. Fig. \ref{fig:SMF_controlled_Atw_effect_MFD} shows the wall normal distribution of the MFD for the controlled configurations relative to the uncontrolled case. It is shown that ahead of the second Mack mode growth ($x<500$) as the amplitude of the streaks increases the near-wall MFD also increases, which is consistent with the downstream stronger stabilization of the boundary layer previously shown in Fig. \ref{fig:SMF_controlled_Cf_tzavg}. Overall, the control method is effective for second Mack mode dominated transition to turbulence scenarios, and optimization strategies may be investigated to grow greater amplitude streaks and enhance the performance of the control method.

\subsubsection{Sensitivity of control method effectiveness to streak wavelength}\label{sec:SMF_ks_effect}
Previous work \citep{Boscagli2025_SCITECH} showed that near-optimum second Mack mode stabilization can be achieved via control-streaks with a wavelength ($\lambda_{z,s}=2\pi/k_s$) that is approximately 8-10 times the boundary layer thickness in the region of the second Mack mode maximum amplification. For this case study this roughly corresponds to $k_s=4$, which is the configuration investigated in the previous sections. For similar hypersonic conditions, previous numerical studies \citep{Paredes2019, Zhou2023} showed transition delay via second Mack mode stabilization can be achieved with a streak wavelength that is approximately of the same size as the optimally growing streak, provided that the amplitude of the streaks is around $As_u=0.1$ (table \ref{tab:review_streaks}). This optimal wavelength would only correspond to a fraction of the size of the local boundary layer thickness. 

\begin{figure}
  \centering 
  \subfloat[]{\includegraphics[width=0.48\textwidth]{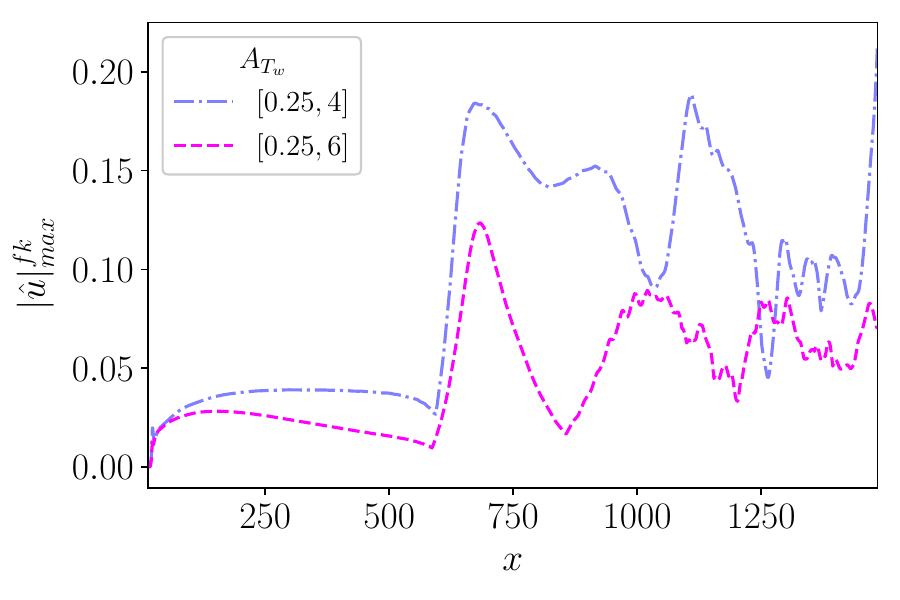}\label{fig:SMF_ks_effect_streaks}}  
\hfill
  \centering 
  \subfloat[]{\includegraphics[width=0.48\textwidth]{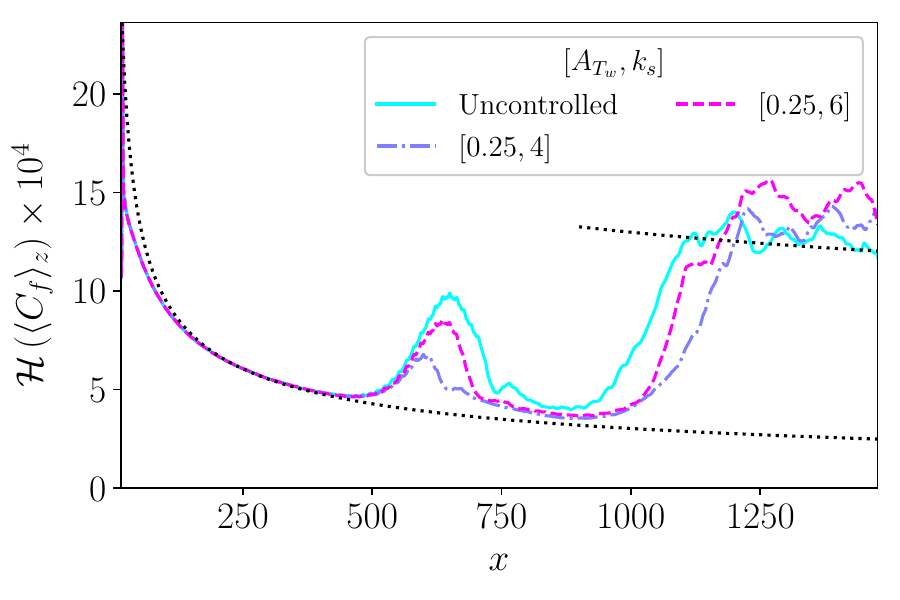}\label{fig:SMF_ks_effect_Cf_tzavg}}
\caption{DNS analysis of the SMF case: (\textit{a}) effect of $k_s$ on the amplitude of the control streaks ($x<500$); (\textit{b}) effect of control method on the streamwise distribution of  the (time) envelope of the spanwise-averaged skin friction coefficient. The grey dotted lines in (\textit{b}) indicate the laminar and turbulent correlations.}
\label{fig:SMF_controlled_ks_effect}
\end{figure}

In this section the sensitivity of the control method effectiveness to changes in streak wavelength is further investigated. Compared to the studies in the previous section, the streak wavenumber is increased to $k_s=6$, with the spanwise temperature variation ($A_{T_w}=0.25$) and spanwise grid resolution ($n_z=128$) held constant. It is found that with a reduction of the streak wavelength from $\lambda_{z,s} \approx 8.7\delta_{99,in}$ ($k_s=4$) to $5.8\delta_{99,in}$ ($k_s=6$), the amplitude of the control streaks reduces along the streamwise direction (Fig. \ref{fig:SMF_ks_effect_streaks}, $100<x<500$), likely due to viscous dissipation effects. Relative to the case with $k_s=4$, the combined reduction of both amplitude and wavelength of the control streaks for the case with $k_s=6$ leads to a decrease in control method effectiveness both in terms of second Mack mode stabilization and transition delay (Fig. \ref{fig:SMF_ks_effect_Cf_tzavg}). This indicates that the generation of large ($As_u>0.1$) amplitude streaks through spanwise non-uniform surface temperature may requires greater energy input for smaller wavelengths. This is further investigated in the next section, where the effect of the control method on transition to turbulence is also assessed for the first Mack mode oblique breakdown scenario.   
 
\subsection{First Mack mode oblique breakdown}\label{sec:transition_FMO}
\begin{figure}
\centering
\includegraphics[trim={0cm 3.0cm 0cm 3.0cm},clip,width=\textwidth]{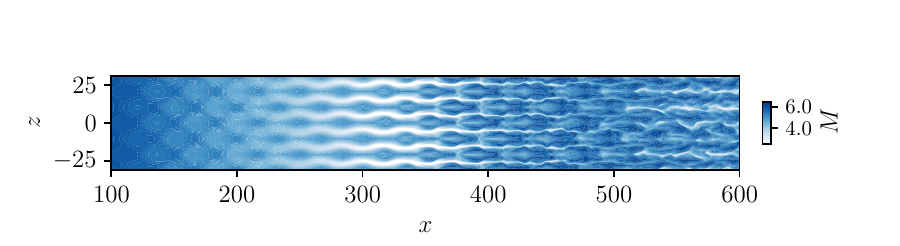}
\caption{DNS analysis of the baseline, uncontrolled FMO case. Instantaneous distribution of streamwise Mach number at $y=2.5$. For better visualization, the figure aspect ratio is distorted by $20\%$, and only $x\in[100,600]$ is shown.}
\label{fig:uncontrolled_FMO_mach}
\end{figure}

\begin{figure}
\centering
\includegraphics[width=0.48\textwidth]{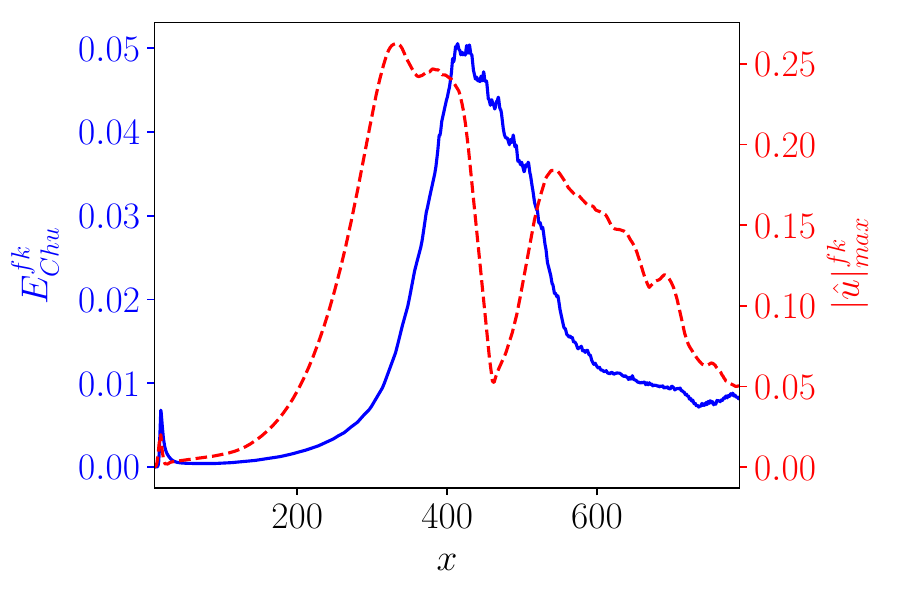}
\caption{DNS analysis of the uncontrolled FMO case. Streamwise distribution of first Mack mode, $(f,k)=(1,3)$, energy (left y-axis, blue solid line) and amplitude of streaks, $(f,k)=(0,6)$, (right y-axis, red dashed line) for the uncontrolled configuration.}
\label{fig:uncontrolled_FMO_streaks_EChu}
\end{figure}

In this section the effect of spanwise non-uniform surface temperature on transition to turbulence is investigated for the first Mack mode oblique (FMO) breakdown scenario. As detailed in section \ref{sec:actuator}, transition to turbulence is promoted by introducing a pair of oblique waves, $(f,k)=(1,\pm 3)$, with a temporal frequency characteristic of the first Mack mode \citep{Franko2013}. Fig. \ref{fig:uncontrolled_FMO_mach} depicts the Mach number distribution at $y=2.5$. The large amplitude oblique waves introduced through the blowing and suction boundary condition ($x_{c,strip}=17.5$) non-linearly interact and generate streaks, $(f,k)=(0,6)$, that rapidly grow ($x\approx 200$) and lead to breakdown to turbulence via multiple non-linear interactions \citep{Franko2013}. The amplitude of the streaks peaks to $|\hat{u}|^{06}_{max} \approx 0.25$ at $x \approx 300$, upstream of the first Mack mode energy peak, $x \approx 400$ (Fig. \ref{fig:uncontrolled_FMO_streaks_EChu}). 
 
Previous numerical studies for supersonic \citep{Paredes2017, Sharma2019} and hypersonic \citep{Ren2016} conditions showed that effective stabilization of first Mack mode and transition delay can be achieved by finite amplitude control streaks ($As_u>0.1$), with a wavenumber that is approximately 4 to 5 times the fundamental wavenumber of the first Mack mode. For the configuration under investigation and based on the non-dimensionalization introduced in section \ref{sec:converegence}, this corresponds to optimal streaks wavenumber for first Mack mode stabilization $k_s=12$. Based upon the studies presented in section \ref{sec:SMF_ks_effect}, it can be anticipated that the generation of strong, closely spaced streaks with spanwise non-uniform surface temperature requires considerable thermal energy of the control streaks. Thus, the effect of weak ($As_u<0.1$) control-streaks on first Mack mode stabilization is investigated for a range of spanwise wavenumber, $k_s$, with peak to peak spanwise temperature variation ($A_{T_w}=0.25$) held constant.     

\begin{figure}
  \centering 
  \subfloat[]{\includegraphics[width=0.48\textwidth]{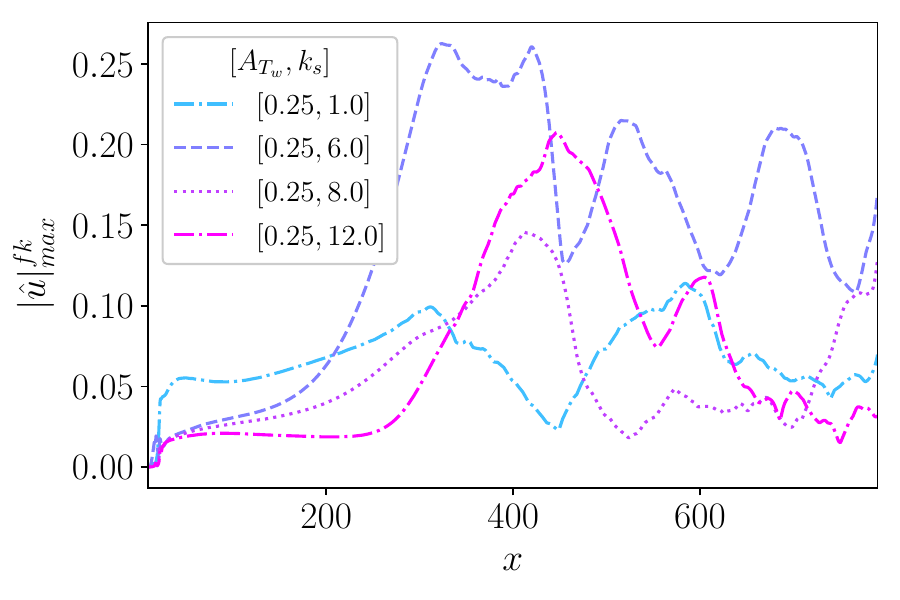}\label{fig:FMO_ks_effect_streaks}}
\hfill
  \centering 
  \subfloat[]{\includegraphics[width=0.48\textwidth]{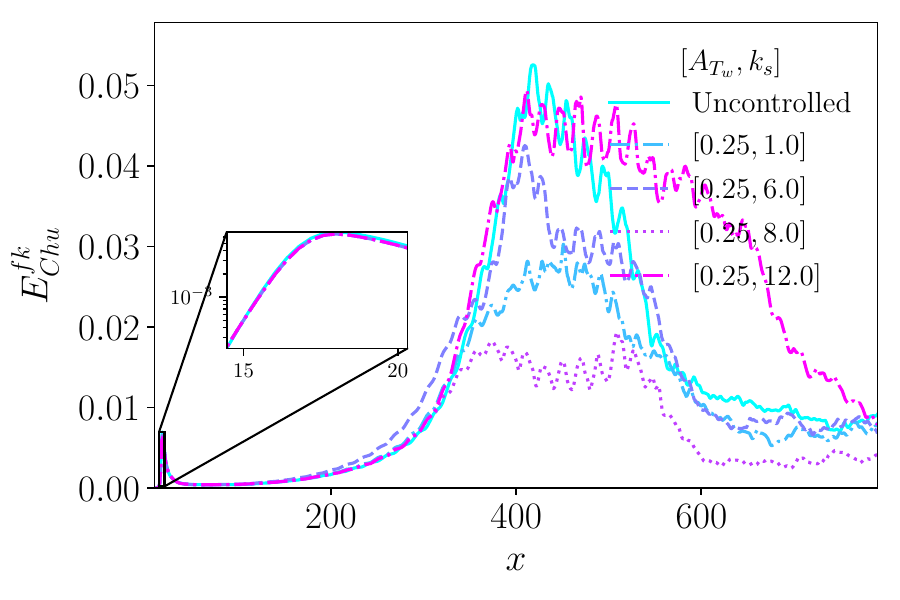}\label{fig:FMO_ks_effect_Echu}}
\caption{DNS analysis of the FMO case; effect of streak wavelength ($k_s$) on the streamwise distribution of (\textit{a}) control-streaks amplitude and (\textit{b}) first Mack mode, $(f,k)=(1,3)$, energy. The inset in (\textit{b}) indicates the modal energy of the forcing disturbance.}
\label{fig:FMO_ks_effect_modal}
\end{figure}

\begin{figure}
\centering
\includegraphics[trim={0cm 0cm 0cm 0.0cm},clip,width=\textwidth]{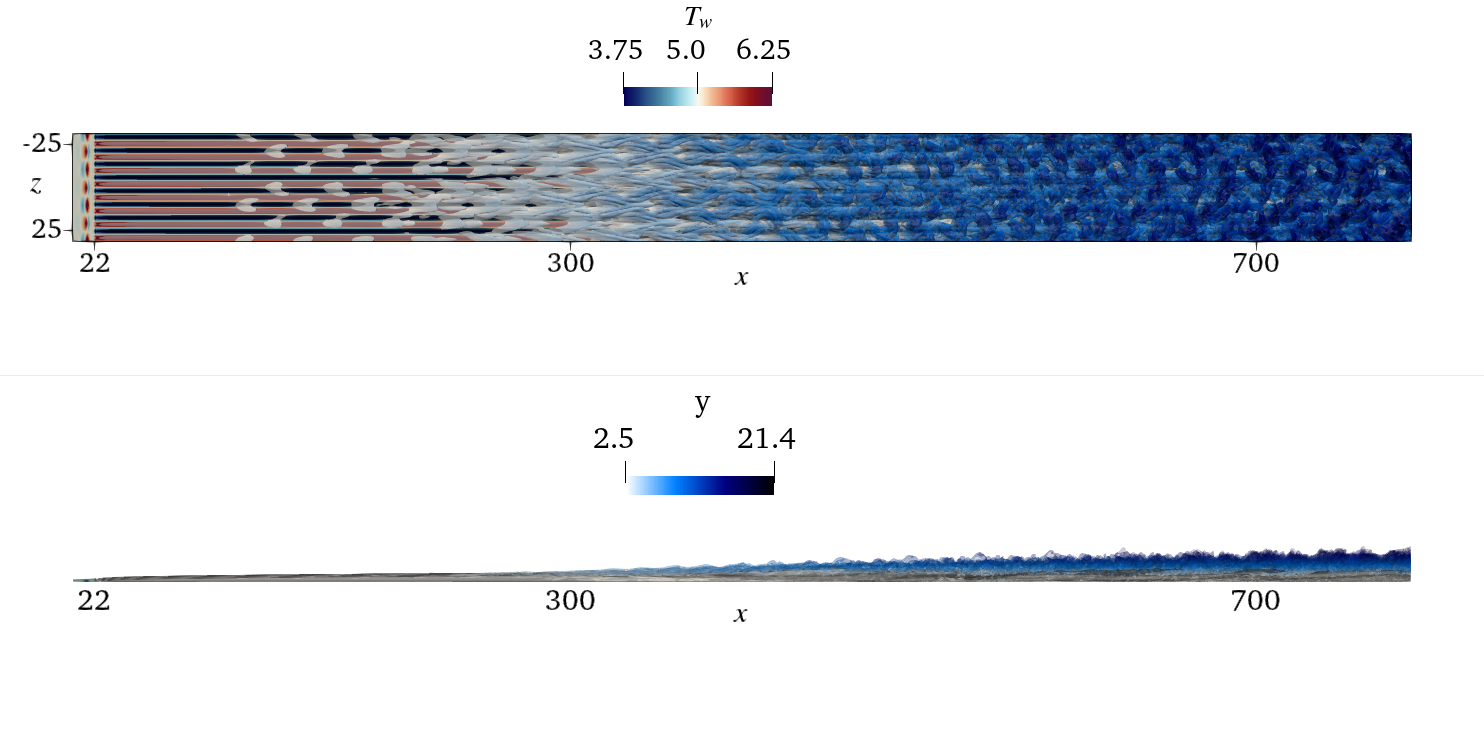}
\caption{DNS analysis of the controlled FMO case. Top (upper) and front (bottom) view of  transition to turbulence via first Mack mode oblique breakdown. $M_{\infty}=6$, $Re_{unit}=7.2 \times 10^6$1/m, $T_{w,base}=5$, $A_{T_w}=0.25$, $k_s=8$. Isosurfaces of Q-criterion ($10^{-8}$) colored with wall normal distance ($y)$. Black and white isosurfaces ($u^{08}=\pm 0.01$) depict instead high and low speed streaks, respectively.}
\label{fig:FMO_controlled}
\end{figure}

\begin{figure}
\centering
\includegraphics[width=0.48\textwidth]{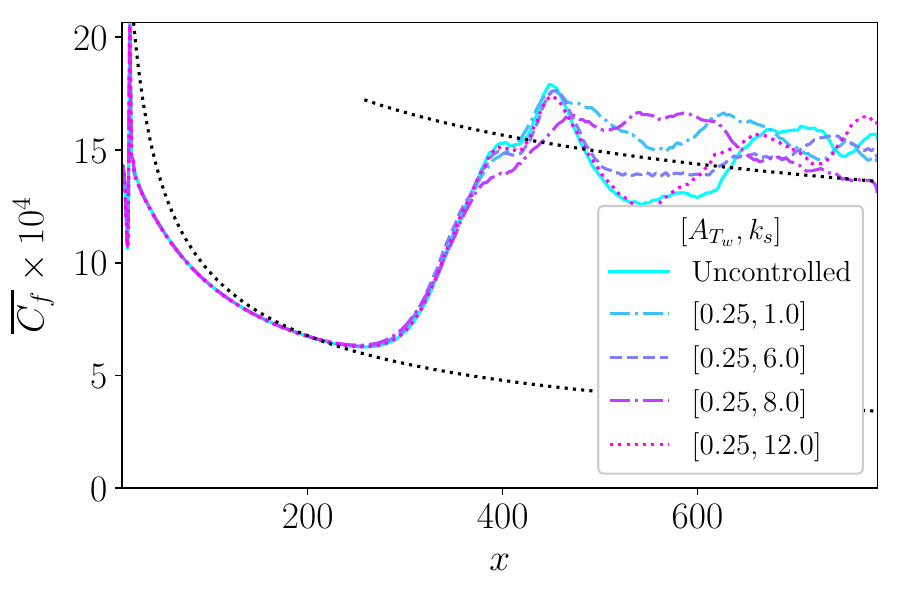}
\caption{DNS analysis of the FMO case; effect of streak wavelength on the streamwise distribution of the time and spanwise-averaged skin friction coefficient. Grey dotted lines indicate the laminar and turbulent correlations.}
\label{fig:FMO_ks_effect_Cf}
\end{figure}

As the wavenumber of the control-streaks is increased from $k_s=1$ to $12$, the amplitude of the control-streaks ($|\hat{u}|^{0k_s}_{max}$) decreases nearly monotonically for $x<200$ (Fig. \ref{fig:FMO_ks_effect_streaks}) from approximately 0.05 to 0.02. For the controlled case with $k_s=6$, at $x\approx 200$ the amplitude of the streaks grows almost exponentially due to the streaks formation associated with the triadic interaction of the first Mack mode waves, which also results in streaks with the same wavenumber $k=6$ (Fig. \ref{fig:uncontrolled_FMO_streaks_EChu}. The peak amplitude of the energy of the first Mack mode ($x\approx 400$ in Fig. \ref{fig:FMO_ks_effect_Echu}) is non-monotonic relative to the change in control-streaks amplitude. For the configuration with $k_s=12$, the first Mack mode is almost not affected by the weak control streaks. A maximum reduction in the first Mack mode energy is achieved for $k_s=8$. Despite this reduction in first Mack mode energy, the flow topology of the controlled configuration (Fig. \ref{fig:FMO_controlled}) is similar to the uncontrolled configuration (Fig. \ref{fig:uncontrolled_FMO_mach}), and none of the configurations investigated affect laminar to turbulent transition (Fig. \ref{fig:FMO_ks_effect_Cf}). This is the combined result of transition being dominated by the rapid growth of streaks due to non-linear interaction of the first Mack mode oblique waves, and the control streaks not being sufficiently strong to achieve an effective stabilization of the latter. 

\subsubsection{Sensitivity of the amplitude of the control-streaks to spanwise temperature variation}
\begin{figure}
  \centering 
  \subfloat[]{\includegraphics[width=0.48\textwidth]{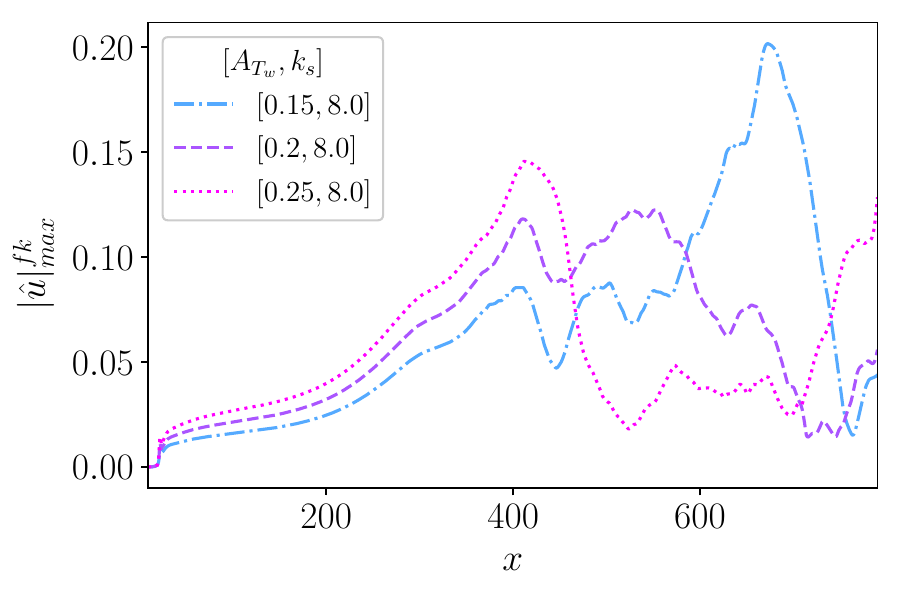}\label{fig:FMO_Atw_effect_streaks}}
\hfill
  \centering 
  \subfloat[]{\includegraphics[width=0.48\textwidth]{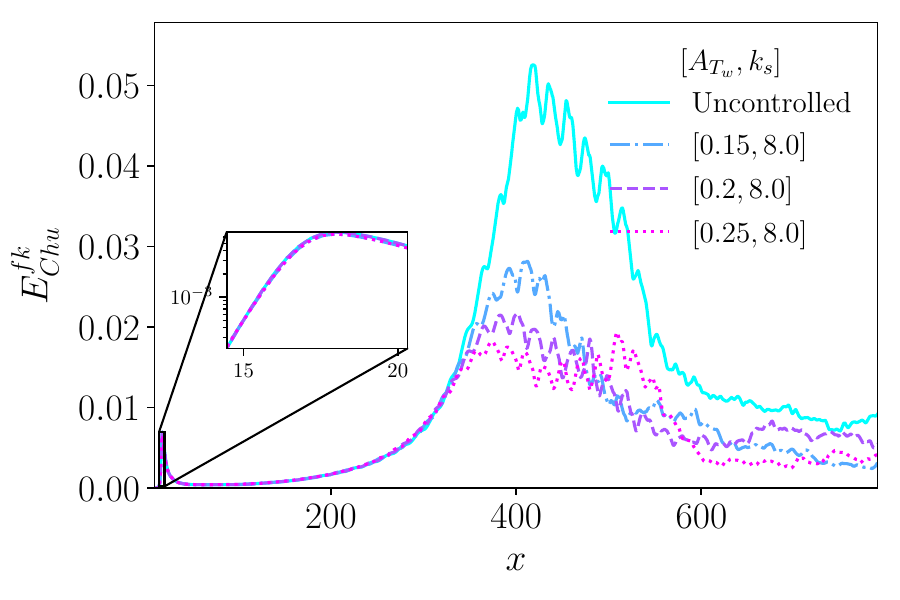}\label{fig:FMO_Atw_effect_Echu}}
\caption{DNS analysis of the uncontrolled FMO case; effect of spanwise temperature variation relative to the base flow ($A_{T_w}$) on the streamwise distribution of (\textit{a}) control-streaks amplitude and (\textit{b}) first Mack mode, $(f,k)=(1,3)$, energy. The inset in (\textit{b}) indicates the modal energy of the forcing disturbance.}
\label{fig:FMO_Atw_effect_modal}
\end{figure}

\begin{figure}
\centering
\includegraphics[width=0.48\textwidth]{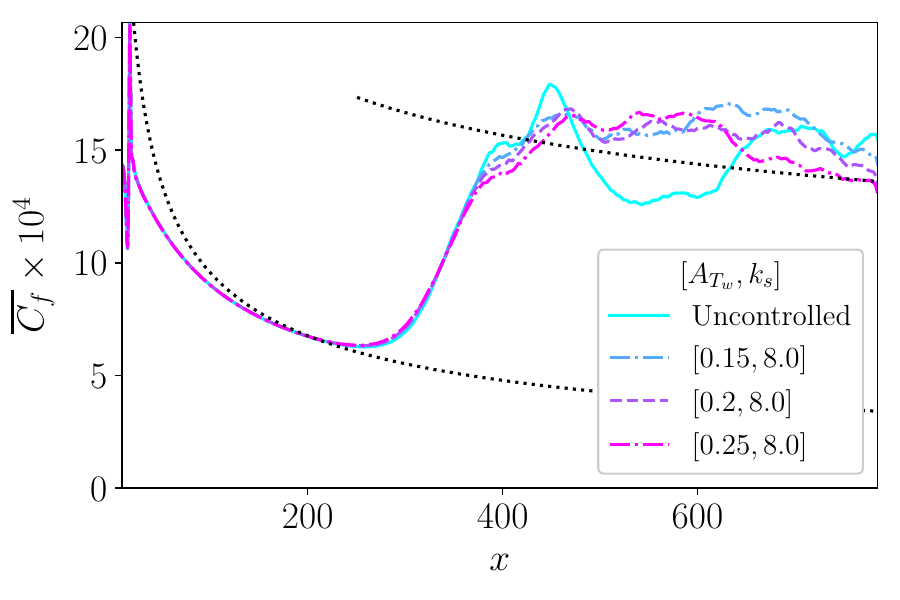}
\caption{DNS analysis of the uncontrolled FMO case; effect of spanwise temperature variation relative to the base flow ($A_{T_w}$) on the streamwise distribution of the time and spanwise-averaged skin friction coefficient. Grey dotted lines indicate the laminar and turbulent correlations.}
\label{fig:FMO_Atw_effect_Cf}
\end{figure}

The previous section showed that weak ($As_u<0.05$) control-streaks are not sufficient to delay transition to turbulence dominated by triadic interaction of first Mack mode oblique waves. For this case study, the maximum reduction in the first Mack mode energy was achieved through the configuration with $k_s=8$. A sensitivity study of the control-streaks to changes in $A_{T_w}$ shows that the amplitude of the control-streaks upstream of transition ($x\approx200$) increases from $|\hat{u}|^{08}_{max} \approx 0.025$ to $0.04$ (Fig. \ref{fig:FMO_Atw_effect_streaks}) as the peak to peak temperature variation is increased from approximately $97\textnormal{K}$ ($A_{T_w}=0.15$) to $162\textnormal{K}$ ($A_{T_w}=0.25$). The increase in streaks reduces the peak amplitude of the first Mack mode energy (Fig. \ref{fig:FMO_Atw_effect_modal}), but overall this is insufficient to significantly affect transition behavior and location (Fig. \ref{fig:FMO_Atw_effect_Cf}). For the combination of Mach number and stagnation enthalpy under investigation, and assuming linear growth of the amplitude of the control streaks with increasing values of $A_{T_w}$ ($A_{T_w} \geq 0.3$), it is estimated that a peak to peak spanwise temperature  variation of more than 420K may be needed to generate control-streaks with $k_s / k_{0} \approx 2.7$ and sufficient amplitude ($As_u \geq 0.1$) to affect transition based on the current literature (table \ref{tab:review_streaks}).

\subsection{Effect of control-streaks on heat transfer}\label{sec:heat_transfer_analysis}
\subsubsection{Mean heat transfer overshoot}
\begin{figure}
  \centering 
  \subfloat[]{\includegraphics[width=0.48\textwidth]{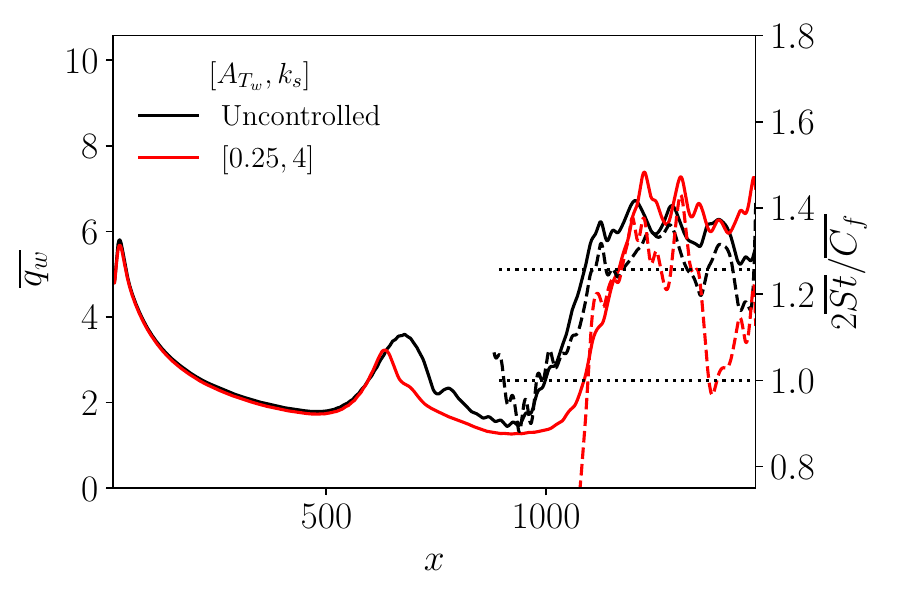}\label{fig:St_number_SMF}}
\hfill
  \centering 
  \subfloat[]{\includegraphics[width=0.48\textwidth]{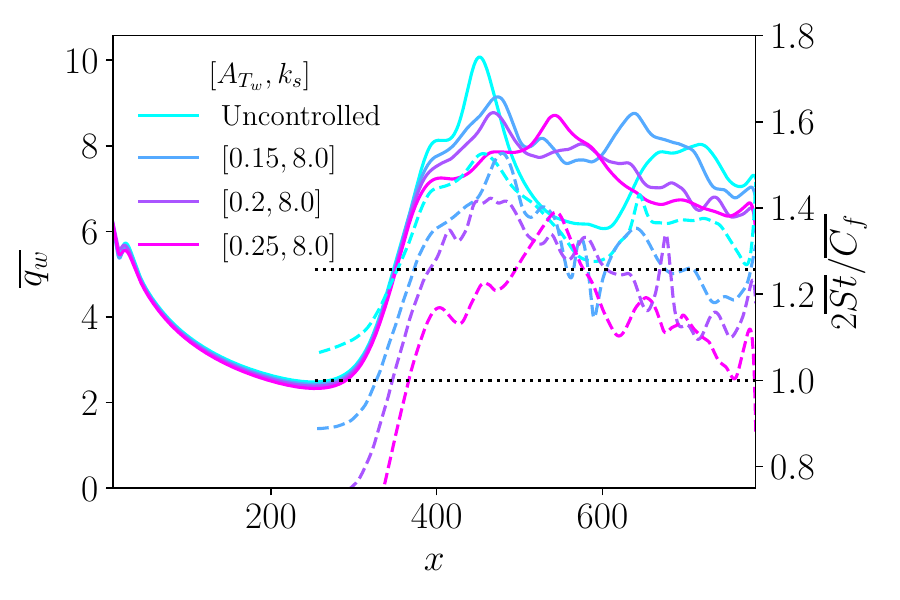}\label{fig:St_number_FMO}}
\caption{Effect of control streaks on heat transfer (solid lines) and skin friction analogy (dashed lines); (\textit{a}) SMF and (\textit{b}) FMO configurations. The black dotted lines indicate the Reynolds ($2\overline{St}/\overline{C_f}=1$) and Chilton-Colburn ($2\overline{St}/\overline{C_f}=Pr^{-2/3}$) analogies.}
\label{fig:St_number}
\end{figure}

In the late stage of laminar to turbulence transition of hypersonic boundary layers, several experimental studies \citep{Malik1990,Horvath2002} reported significant heat transfer (Stanton number, $St$) peaks, which manifest as an overshoot when compared to the prediction based on the Reynolds analogy with skin friction coefficient, $2\overline{St}/\overline{C_f}=1$. \citet{Franko2013} numerically showed that for a Mach 6 boundary layer over a nearly adiabatic ($T_w=6.5$) flat plate with zero pressure gradient, the heat transfer overshoot can be nearly twice the prediction. The peak was observed for first Mack mode oblique (FMO) breakdown, but not for the second Mack mode fundamental resonance (SMF), and it was likely associated to a peak in the wall-normal component of the turbulent heat flux ($\overline{\rho v^{''} T^{''}}$). Local heat-transfer peaks, sometimes referred to as hot spots, significantly affect the design of hypersonic vehicles due to both mean and fluctuating aerothermodynamic loadings \cite{Knight2018}. Thus, it is useful to evaluate the effect of the control-streaks on heat transfer overshoot for both the SMF and FMO configurations, to understand if the overshoot is affected by the control method. The non-dimensional heat transfer at the wall is evaluated based on the time and spanwise averaged Stanton number, which is defined as

\begin{equation}
\overline{St} = \frac{\overline{\left( \mu \frac{\partial T}{\partial y} \right)_w }}{Re_{\infty} M_{\infty} Pr_{\infty} (T_{aw} - \overline{T}_w)} = \frac{\overline{q}_w}{Re_{\infty} M_{\infty} Pr_{\infty} (T_{aw} - \overline{T}_w)}
\end{equation}
where $T_{aw}$ is the adiabatic wall temperature based on the turbulent value of the recovery factor ($r=Pr^{1/3}$), as the objective of this analysis is to evaluate the heat transfer overshoot from the comparison with the skin friction coefficient analogy in the turbulent regime. For the second Mack mode fundamental resonance, the Stanton number in the final stage of transition to turbulence is close to the prediction based on Chilton-Colburn analogy (Fig. \ref{fig:St_number_SMF}), and in agreement with previous studies \citep{Franko2013} there is no significant overshoot (approximately below $1.4$). For the SMF case, the weak control-streaks do not affect the magnitude of the heat transfer overshoot, but only the streamwise location as a result of the delayed laminar to turbulence transition. For the first Mack mode oblique breakdown, the uncontrolled configuration shows heat transfer overshoot approximately 1.65 times greater than the Reynolds analogy prediction (Fig. \ref{fig:St_number_FMO}). For this configuration the control streaks with $k_s=8$ and $A_{T_w}=0.25$ reduce the overshoot down to 1.4, and the benefit increases as $A_{T_w}$ is increased from 0.15, which provides nearly no benefits, to 0.2 and 0.25. Thus, the mechanisms driving this reduction in heat transfer overshoot due to weak control-streaks for the FMO case are further investigated.  

\begin{figure}
  \centering 
  \includegraphics[width=0.8\textwidth]{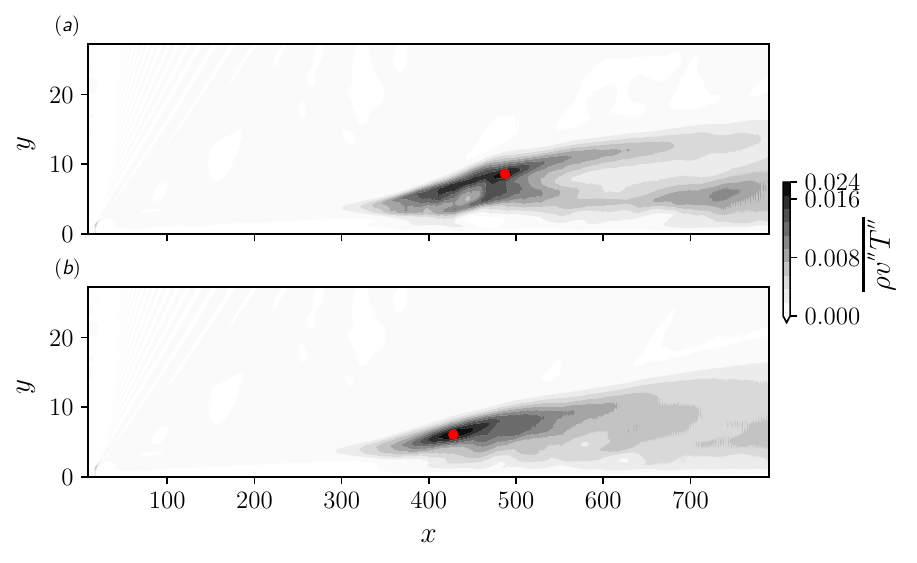}
\caption{DNS analysis of the FMO case; effect of control-streaks on spanwise averaged, wall-normal component of the turbulent heat transfer for the FMO case. (\textit{a}) Uncontrolled and (\textit{b}) controlled ($[A_{T_w},k_s]=[0.25,8]$) configurations. The red marker indicates the location of the positive peak. For better visualization, the figure aspect ratio is set to 8.}
\label{fig:FMO_qheat_V_FA}
\end{figure}

Positive values of the wall-normal component of the turbulent heat flux indicate that high temperature flow is moved away from the wall and low temperature flow is moved towards the wall, and the thermal energy transfer is responsible for the peak in wall heat transfer as previously observed by \citet{Franko2013}. The streamwise and wall-normal distribution of the Favre and spanwise averaged wall-normal component of the turbulent heat flux for the uncontrolled FMO configuration indicates that there is a peak (red marker in Fig. \ref{fig:FMO_qheat_V_FA}a) at approximately the same location of the heat transfer overshoot (Fig. \ref{fig:St_number_FMO}), in agreement with previous studies \citep{Franko2013}. For the controlled configuration the peak in $\overline{\rho v^{''}T^{''}}$ ($x \approx 450$ in Fig. \ref{fig:FMO_qheat_V_FA}b) occurs further upstream compared to the streamwise location of the wall heat transfer overshoot ($x \approx 550$ in Fig. \ref{fig:St_number_FMO}). Thus, the budget of the turbulent kinetic energy (equation \eqref{eq:FA_TKE_budget}) is therefore further inspected.

\begin{figure}
\centering
\includegraphics[width=0.48\textwidth]{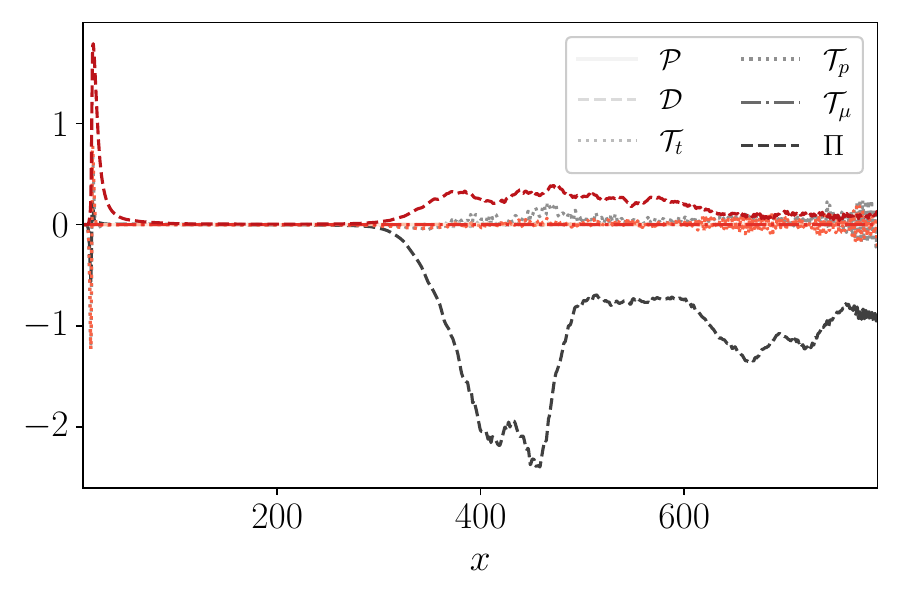}
\caption{DNS analysis of the FMO case; effect of control-streaks on streamwise distribution of spanwise and wall normal integrated turbulent kinetic energy budget terms for the FMO case. Black lines: uncontrolled configuration; red lines: controlled ($[A_{T_w},k_s]=[0.25,8]$) configuration.}
\label{fig:FMO_TKE_budget}
\end{figure}

\begin{figure}
  \centering 
  \includegraphics[clip,width=0.8\textwidth]{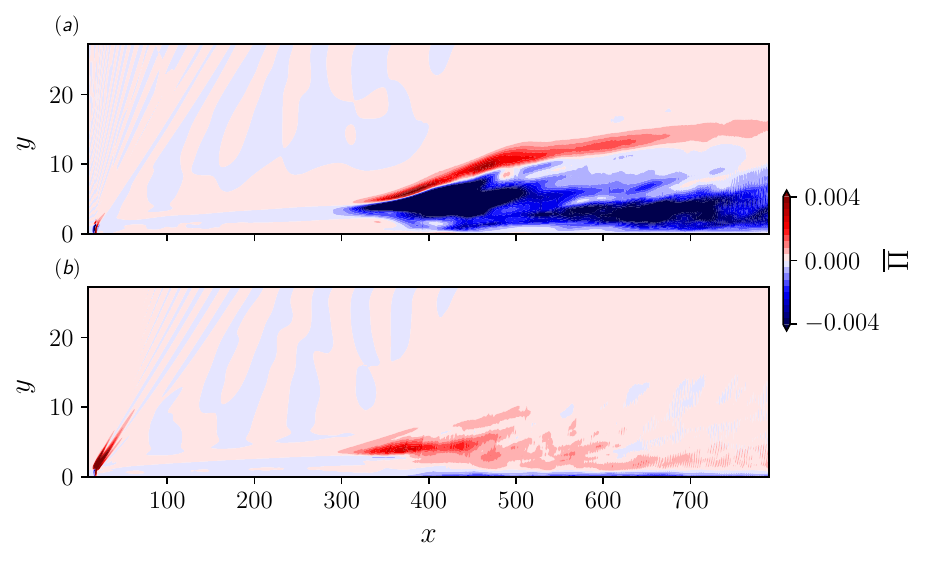}
\caption{DNS analysis of the uncontrolled FMO case; effect of control streaks on spanwise averaged sum of pressure work and pressure dilatation of turbulent kinetic energy for the FMO case. (\textit{a}) Uncontrolled and (\textit{b}) controlled ($[A_{T_w},k_s]=[0.25,8]$) configurations, respectively. For better visualization, the figure aspect ratio is set to 8.}
\label{fig:FMO_TKE_Pi}
\end{figure}

Following \citet{Wilcox1998}, the production term ($\mathcal{P}$) and the pressure-correlation contribution ($\Pi \equiv \Pi_w+\Pi_d$) represent transfer between mean kinetic energy and turbulent kinetic energy, whereas the destruction term ($\mathcal{D}$) represents transfer from turbulent kinetic energy to internal energy. By contrast, the molecular, turbulent, and pressure transport terms ($\mathcal{T}_{\mu}$, $\mathcal{T}_t$, and $\mathcal{T}_p$) redistribute turbulent kinetic energy in space. The streamwise distribution of the spanwise and wall-normal integral of the turbulent kinetic energy budget terms in Fig. \ref{fig:FMO_TKE_budget} indicates that the sum of the pressure work and dilatation ($\Pi$) has the largest contribution, and different polarity for the uncontrolled and controlled configurations after the onset of transition ($x>300$). Positive $\Pi$ indicates a local transfer from mean to turbulent kinetic energy, and vice versa for negative $\Pi$. Relative to the uncontrolled case, the spanwise averaged distribution of $\Pi$ in Fig. \ref{fig:FMO_TKE_Pi} for the controlled configuration shows a reduced negative region, which indicates a reduction in energy transfer from turbulent to mean kinetic energy. The reduction in transfer of turbulent to mean kinetic energy due to the control-streaks may be the driving mechanism behind the reduction in heat transfer overshoot for the mean flow, as a result of a reduced mean kinetic energy to be dissipated at the wall.

\subsubsection{High-frequency heat transfer peak}
\begin{figure}
\centering
\includegraphics[width=0.48\textwidth]{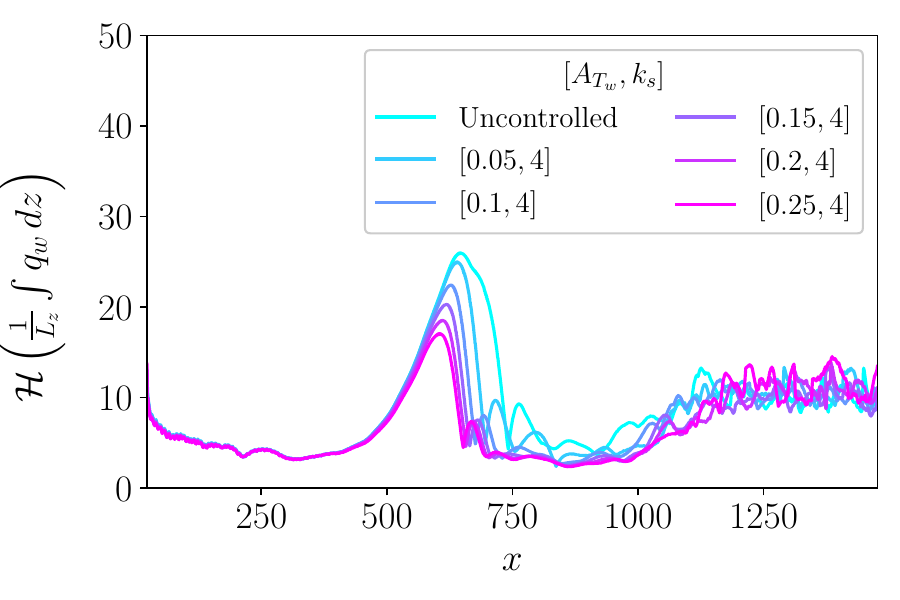}
\caption{DNS analysis of the SMF case; effect of control method on the streamwise distribution of the (time) envelope of the (spanwise-integrated) heat flux.}
\label{fig:SMF_high_frequency_q}
\end{figure}

\begin{figure}
  \centering 
  \subfloat[]{\includegraphics[width=0.48\textwidth]{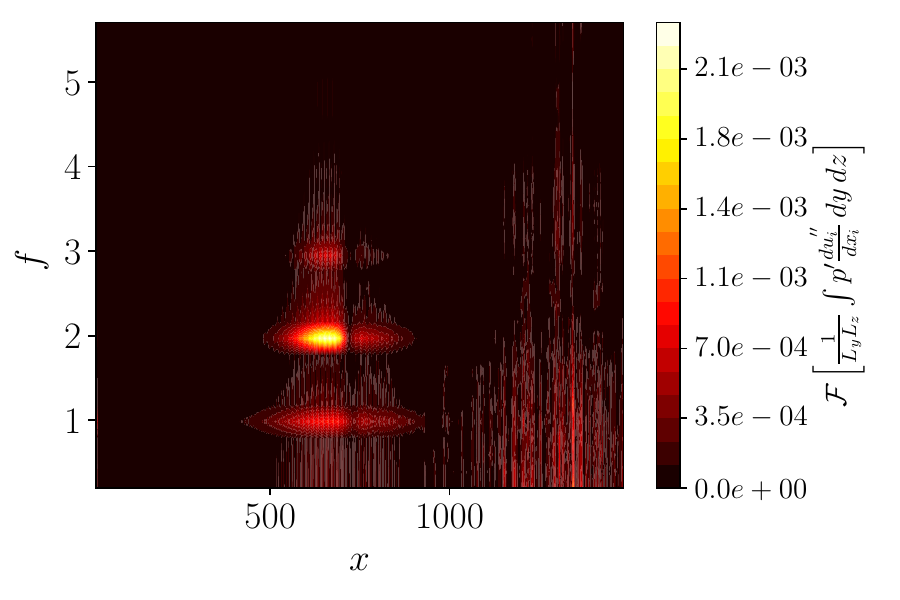}\label{fig:SMF_dilatation_uncontrolled_a}}
\hfill
  \subfloat[]{\includegraphics[width=0.48\textwidth]{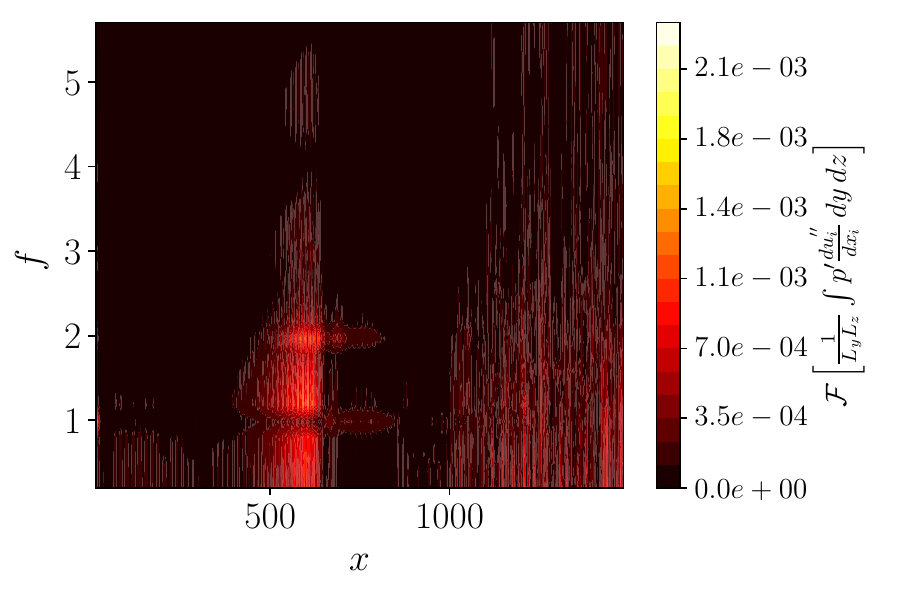}\label{fig:SMF_dilatation_uncontrolled_b}}
\caption{DNS analysis of the SMF case. Streamwise distribution of the temporal Fourier transform of the spanwise and wall-normal integrated dilatation work of turbulent kinetic energy. (\textit{a}) Uncontrolled and (\textit{b}) controlled, $[A_{T_w},k_s]=[0.25,4]$, configurations.}
\label{fig:SMF_dilatation_Fourier}
\end{figure}

\citet{Zhu2018} experimentally showed that a second peak in surface-temperature rise exists, which corresponds to the location of maximum amplification of the second Mack mode. The aerodynamic heating mechanism is attributed to high-frequency dilatation work due to the second Mack mode. The Hilbert-transform of the wall heat-flux integrated along the span depicted in Fig. \ref{fig:SMF_high_frequency_q} retains the high-frequency effect of the second Mack mode on the wall heat transfer ($500<x<900$), and indicates a significant reduction of the amplitude of the hot-spot due to the control streaks. Compared to the uncontrolled configuration, increasing the spanwise temperature variation from $A_{T_w}=0.05$ to $0.25$ leads to a reduction in the high-frequency (spanwise integrated) heat-transfer peaks from approximately 4\% to 34\%. It is acknowledged that as the wall temperature is sinusoidally varied relative to the base flow, the integrated surface heat flux between controlled and uncontrolled configurations is not the same. In the laminar region, this leads to approximately a 3.5\% cooling effect for the case with $A_{T_w}=0.25$ compared to the uncontrolled case ($A_{T_w}=0$). This is significantly lower than the reduction in high-frequency peak heat-flux, which is therefore attributed to the control streaks. The analysis of the budget of the perturbation energy indicates that in the region of second Mack mode maximum amplitude ($500<x<900$), the pressure dilatation work dominates. Relative to the uncontrolled case, the control method reduces the amplitude of the high-frequency dilatation work at the fundamental, first and second super-harmonics of the disturbance forcing angular frequency (Fig. \ref{fig:SMF_dilatation_Fourier}).

\section{Conclusions}\label{sec:conclusions}
For a transitional, Mach 6 boundary layer over a flat plate, this work determined and quantified the effectiveness of control-streaks generated through spanwise non-uniform surface temperature to stabilize first and second Mack mode, and the effect on transition to turbulence. For the second Mack mode fundamental resonance scenario it is shown that the control-streaks stabilize the second Mack mode, reducing the high-frequency (second Mack mode related) peaks in the skin friction coefficient. For control-streaks with an amplitude of approximately 4\% of the freestream speed this reduction can be up to 30\% relative to the uncontrolled configuration. Based on conventional streak amplitude definition and the existing literature, the control-streaks under investigation are classified as weak streaks. The weak control-streaks can delay transition to turbulence and for the range of configurations investigated the delay is approximately 40 to 60 times the local boundary layer thickness. The control streaks primarily produce a beneficial modification of the spanwise-mean flow, and the resulting transition delay is largely insensitive to the relative spanwise phase investigated.

For the first Mack mode oblique breakdown, optimal (based on the existing literature) and non-optimal spanwise wavelengths of the control-streak have been investigated. Overall, the work showed that the weak optimal streaks can stabilize the first Mack mode, but there is no effect on transition to turbulence. Relative to the configuration with control-streaks wavelength approximately 3 times the fundamental wavelength of the first Mack mode, it is shown that, for the combination of Mach number and stagnation enthalpy investigated, and based on a linear scaling of the control-streaks amplitude with wall temperature spanwise variation, approximately more than 420K temperature difference between the hot and cold patches may be needed to affect transition to turbulence. 

To the authors' knowledge, the effect of control streaks on mean and fluctuating heat-transfer peaks has never been reported before, and it is a novel contribution of this work, specifically for weak control-streaks generated through thermal modulation. For the second Mack mode fundamental resonance, the control-streaks with approximately 3.5\% amplitude are able to reduce the maximum amplitude of high-frequency heat transfer fluctuations by 34\% relative to the uncontrolled configuration. In agreement with previous studies, there is no overshoot for second Mack mode dominated breakdown to turbulence, and this is not affected by the control-streaks. For first Mack mode oblique breakdown, it is found that control-streaks with relatively low amplitude (below 5\%) are able to reduce the overshoot by up to 15\%. 

Overall, these results extend previous findings by showing how spatially non-uniform thermal conditions can delay transition to turbulence and reduce the amplitude of mean and high-frequency hot spots. The insights gained here provide guidelines for further optimization of this control strategy and its integration into future hypersonic vehicle designs.    
  
\begin{acknowledgments}
This research received financial support of Dstl through the WSRF program (task number 0105). The authors acknowledge EPSRC for the computational time made available on the UK supercomputing facility ARCHER2 via the UK Turbulence Consortium (EP/R029326/1), and computational resources and support provided by the Imperial College Research Computing Service (\url{http://doi.org/10.14469/hpc/2232}).
\end{acknowledgments}

\appendix

\section{Scale resolution, rarefaction and statistical convergence}
\label{app:verification}

This appendix summarizes additional verification diagnostics used to support the DNS analysis: grid resolution in Kolmogorov units, rarefaction assessment, and statistical convergence of heat-transfer peaks, turbulent statistics and TKE-budget terms. Two representative (controlled) configurations have been used, with non-dimensional amplitude of the spanwise temperature variation set to $A_{T_w}=0.25$. For conciseness, in the sections  below these are simply referred to as SMF and FMO.

\subsection{Grid resolution in Kolmogorov units}
\label{app:grid_eta}

\begin{figure}
\centering
\includegraphics[width=0.48\textwidth]{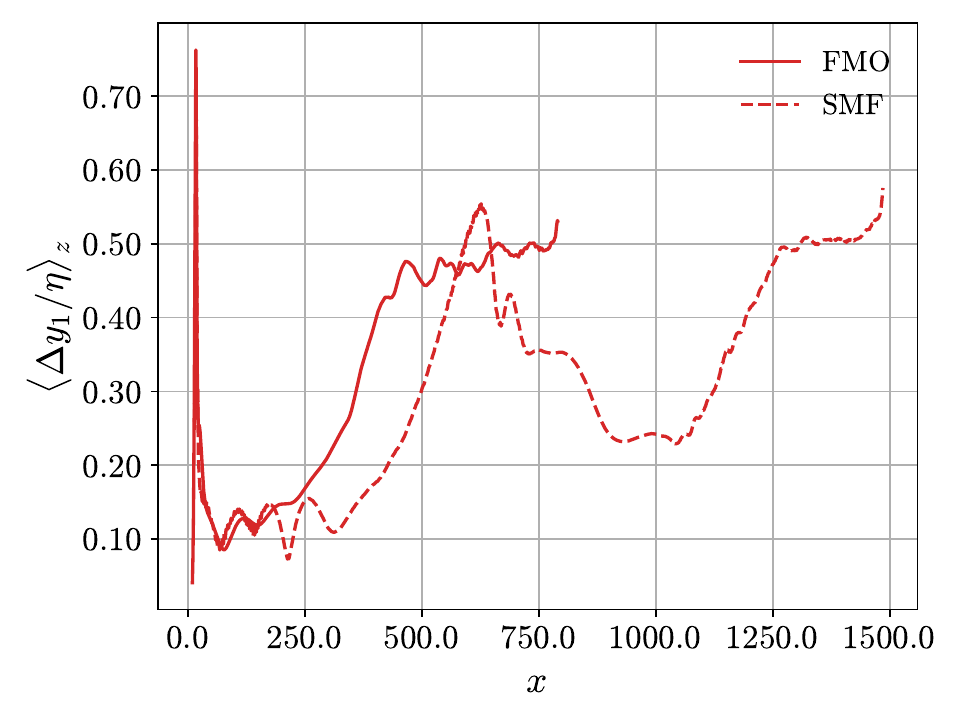}
\caption{Streamwise distribution of (spanwise-averaged) first off-wall spacing to Kolmogorov length scale ratio.}
\label{fig:app_dy1_x}
\end{figure}

The grid resolution is quantified relative to the Kolmogorov length scale (eq.~\ref{eq:kolmogorov_length scale}) through
\begin{equation}
\frac{\Delta x}{\eta}, \qquad
\frac{\Delta y}{\eta}, \qquad
\frac{\Delta z}{\eta},
\end{equation}
with additional focus on first off-wall spacing
\begin{equation}
\frac{\Delta y_1}{\eta}.
\end{equation}
Figure~\ref{fig:app_dy1_x} shows $\langle \Delta y_1/\eta \rangle_z$ as a function of $x$, and table~\ref{tab:app_grid_eta} reports the resolution over the whole computational domain. The initial peak in figure~\ref{fig:app_dy1_x} corresponds to the disturbance forcing region ($x\in[15,20]$).

\begin{table}[t]
\centering
\caption{Range of grid spacing in Kolmogorov units over the computational domain.}
\label{tab:app_grid_eta}
\begin{tabular}{lcccc}
\hline
Case & $\Delta x/\eta$ & $\Delta y/\eta$ & $\Delta z/\eta$ & $\Delta y_1/\eta$\\
\hline
FMO & [1.00e-04,\,1.38e+01] & [1.00e-04,\,6.84e+00] & [1.00e-04,\,8.69e+00] & [2.18e-02,\,8.67e-01]\\
SMF & [1.00e-04,\,2.12e+01] & [1.00e-04,\,1.02e+01] & [1.00e-04,\,1.42e+01] & [5.10e-02,\,9.13e-01]\\
\hline
\end{tabular}
\end{table}

\subsection{Rarefaction assessment}
\label{app:rarefaction}

\begin{figure}
  \centering 
  \subfloat[]{\includegraphics[trim={0.5cm 3cm 0.5cm 3cm},clip,width=0.6\textwidth]{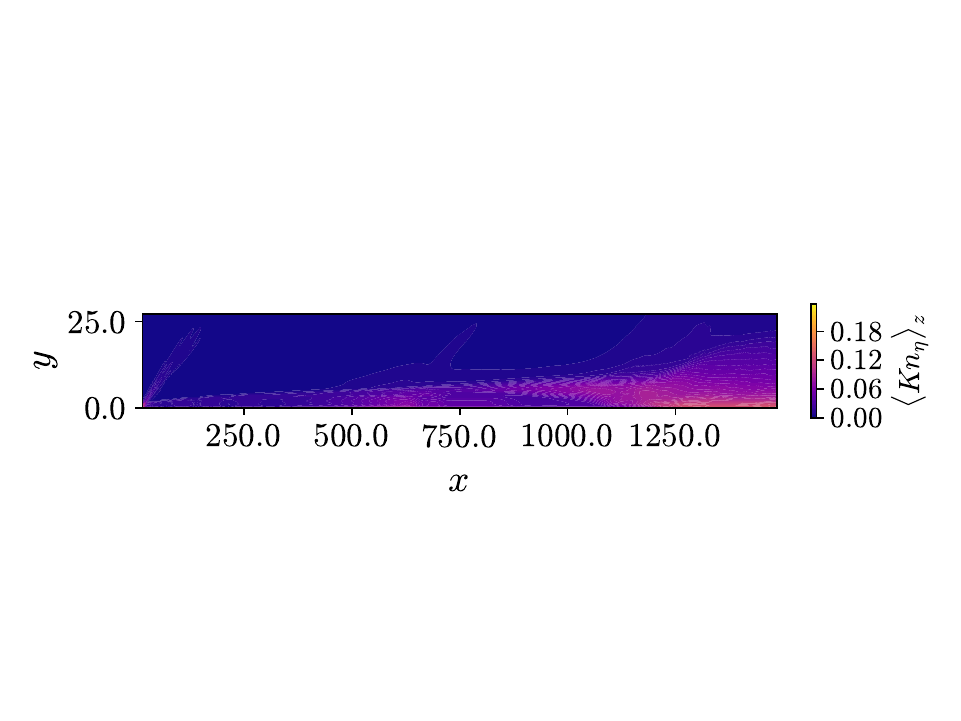}\label{fig:SMF_Kn}}
\hfill
  \centering 
  \subfloat[]{\includegraphics[trim={0.5cm 3cm 0.5cm 3cm},clip,width=0.6\textwidth]{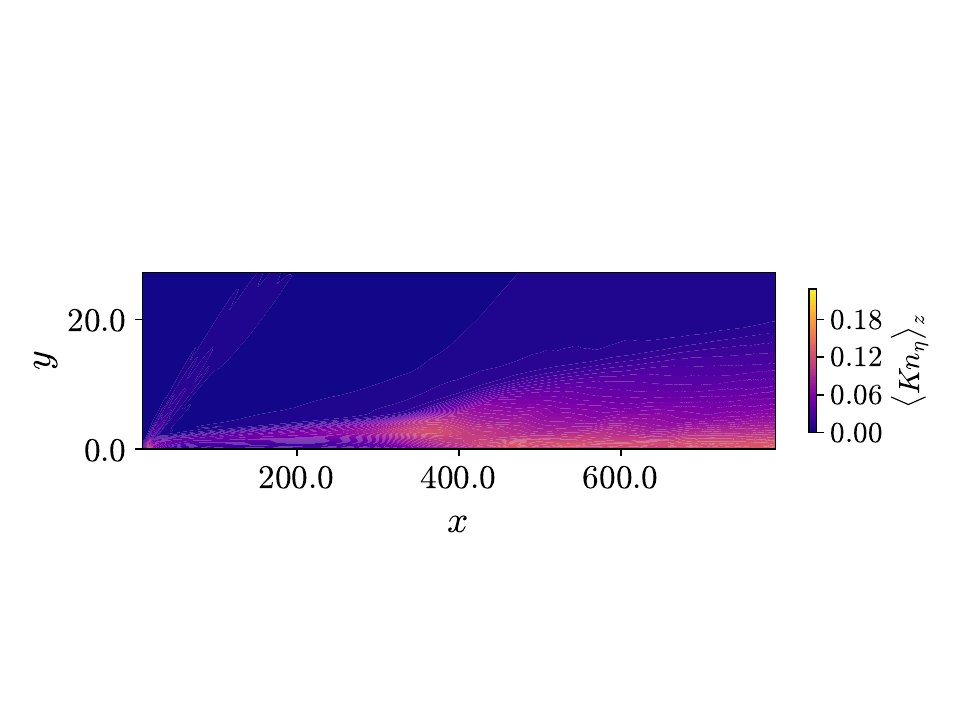}\label{fig:FMO_Kn}}
\caption{DNS analysis of spanwise averaged distribution of Knudsen number for (\textit{a}) SMF and (\textit{b}) FMO configurations. Figure aspect ratio set to 8 for better visualization.}
\label{fig:app_kn_map}
\end{figure}

Rarefaction effects are assessed using the Knudsen number ($Kn_{\eta}$) based on the Kolmogorov length scale, as defined in equation \ref{eq:knudsen}. Figure~\ref{fig:app_kn_map} shows the meridional distribution of  $\langle Kn_{\eta}\rangle_z$, and Table~\ref{tab:app_kn_stats} reports global statistics. The maximum local Knudsen number based on the Kolmogorov length scale is approximately $\mathrm{Kn}_\eta = \lambda/\eta \approx 0.29$, occurring only within a narrow region adjacent to the wall, while the global Knudsen number remains $Kn\approx2.03\times 10^{-3}$. Previous Direct Simulation Monte Carlo (DSMC)--DNS comparisons \citep{Gallis2021} of compressible homogeneous isotropic turbulence reported comparable values of $\mathrm{Kn}_\eta \approx 0.2$ at similar global Knudsen number ($\mathrm{Kn} \approx 10^{-3}$), yet found that the turbulence energy cascade and statistical properties remained essentially unchanged between the kinetic and continuum descriptions. Although the present flow is fundamentally different, these results provide supporting evidence that the localized values of $\mathrm{Kn}_\eta$ encountered here are unlikely to introduce significant non-continuum effects on the resolved turbulence statistics.

\begin{table}[t]
\centering
\caption{Statistics of $Kn_{\eta}$ over the computational domain.}
\label{tab:app_kn_stats}
\begin{tabular}{lcccc}
\hline
Case & Min & Mean & 95th percentile & Max\\
\hline
FMO & 9.44e-09 & 4.08e-02 & 1.26e-01 & 2.87e-01\\
SMF & 8.80e-09 & 2.90e-02 & 1.04e-01 & 2.90e-01\\
\hline
\end{tabular}
\end{table}

\subsection{Statistical convergence of heat transfer peak, turbulent statistics and TKE-budget terms}
\label{app:convergence}

\begin{figure}
\centering
\includegraphics[width=0.48\textwidth]{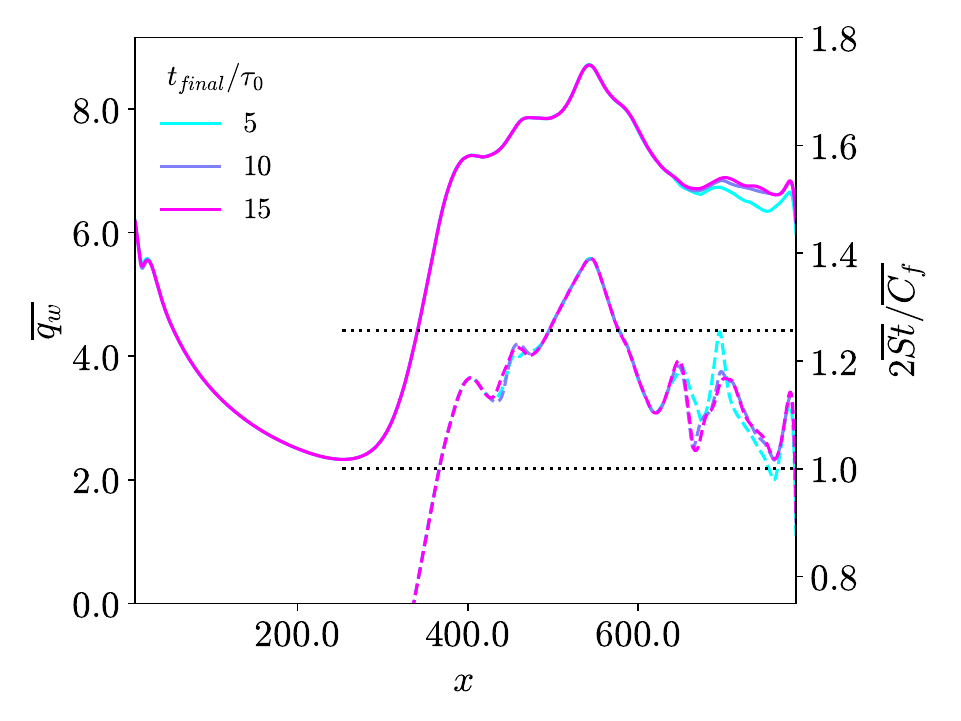}
\caption{DNS analysis of the effect of averaging-window length on heat transfer (solid lines) and skin friction analogy (dashed lines) for the FMO configuration. The black dotted lines indicate the Reynolds ($2\overline{St}/\overline{C_f}=1$) and Chilton-Colburn ($2\overline{St}/\overline{C_f}=Pr^{-2/3}$) analogies.}
\label{fig:qheat_convergence_FMO}
\end{figure}

For the FMO case the effect of averaging-window length ($t_{final}$) on time and spanwise averaged heat transfer coefficient is monitored to assess statistical convergence. Over a time horizon of $t_{final}=15\tau_0$ (similar to \citet{Franko2014}), the heat transfer overshoot remains unchanged as $t_{final}$ is progressively increased from $5\tau_0$ to $15\tau_0$ (Fig. \ref{fig:qheat_convergence_FMO}) with only minor variation in the fully turbulent regime region ($x>700$). 

\begin{figure}
\centering
\includegraphics[width=0.48\textwidth]{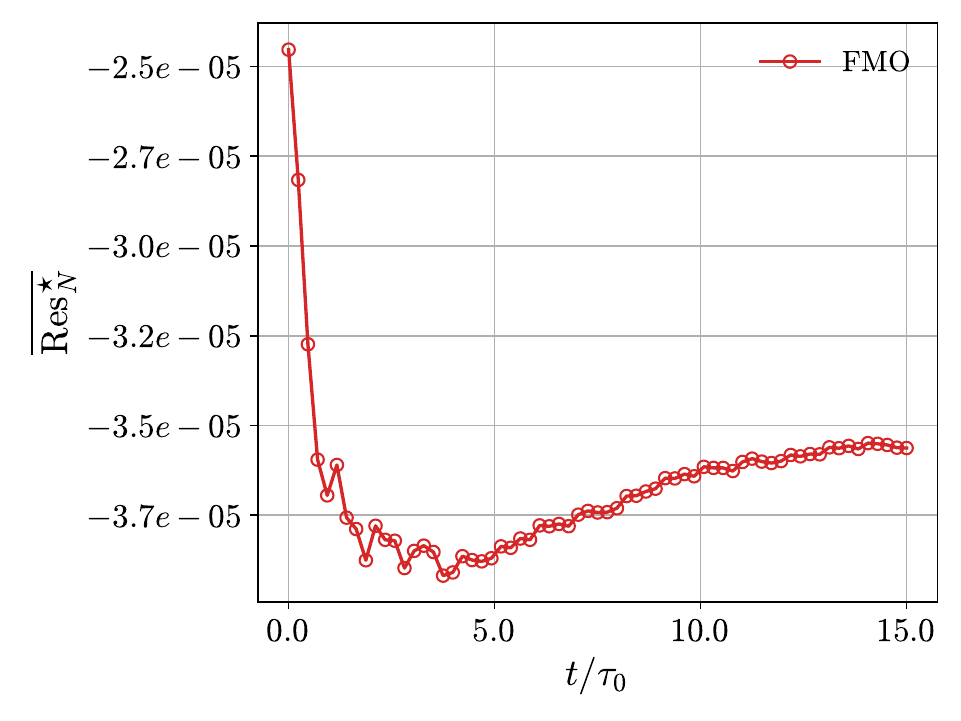}
\caption{DNS analysis of running-means of normalized residuals convergence for the FMO configuration.}
\label{fig:RES}
\end{figure}

\begin{figure}
\centering
\includegraphics[width=0.6\textwidth]{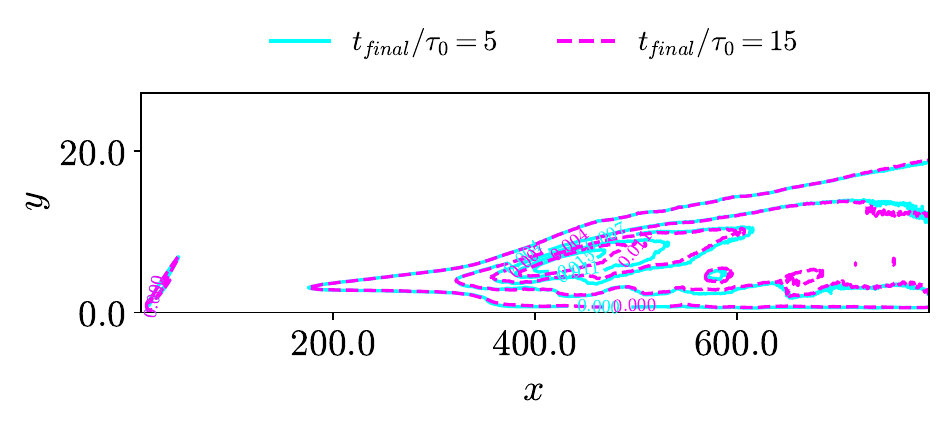}
\caption{DNS analysis of the effect of averaging-window length on spanwise-averaged turbulent heat flux distribution for the FMO configuration.}
\label{fig:qheat_turb_convergence}
\end{figure}

For statistical convergence diagnostics of turbulent statistics, the instantaneous residual of the turbulent kinetic energy budget is assembled as
\begin{equation}
\mathrm{Res}
\equiv
\sum_i \mathcal{T}_i
=
\frac{\partial \left(\langle \rho \rangle \kappa\right)}{\partial t}
+
\langle \rho \rangle \breve{u}_j \frac{\partial \kappa}{\partial x_j}
-
\mathcal{P}
+
\mathcal{D}
-
\mathcal{T}_{t}
-
\mathcal{T}_{p}
-
\mathcal{T}_{\mu}
-
\Pi_{w}
-
\Pi_{d}.
\end{equation}
Statistical convergence is evaluated from running means of volume-averaged ($\left\langle \cdot \right\rangle_V$) terms:
\begin{equation}
\overline{\mathcal{T}_{i,N}}
=
\frac{1}{N}\sum_{n=1}^{N}
\left\langle \mathcal{T}_i \right\rangle_V^{(n)},
\qquad
\overline{\mathrm{Res}_N}
=
\frac{1}{N}\sum_{n=1}^{N}
\left\langle \mathrm{Res} \right\rangle_V^{(n)}.
\end{equation}
The normalized residual is defined as
\begin{equation}
\overline{\mathrm{Res}_N^{\star}}
=
\frac{\left|\overline{\mathrm{Res}_N}\right|}
{\max_i\left|\overline{\mathcal{T}_{i,N}}\right|},
\end{equation}
where $N$ is the averaging-window length. Figure~\ref{fig:RES} shows residuals convergence for the FMO configuration. For $t_{final}=15\tau_0$, the normalized convergence approaches a plateau and increasing $t_{final}$ from $5\tau_0$ to $15\tau_0$ the normalized residuals change by less than $8\%$, with no noticeable changes in either turbulent kinetic energy budget terms nor in turbulence heat flux distribution and peaks (Fig.~\ref{fig:qheat_turb_convergence}). Overall, for the scope of this work, which is focused on the analysis of the transition process, $t_{final}=5\tau_0$ provides sufficiently converged statistics, and a further increase in $t_{final}$ does not provide qualitative nor quantitative changes in the main conclusions of the DNS studies.

\section{Assessment of control method effect on the spanwise symmetry of the flow}\label{app:symmetry}
\begin{figure}
  \centering 
  \subfloat[]{\includegraphics[width=0.48\textwidth]{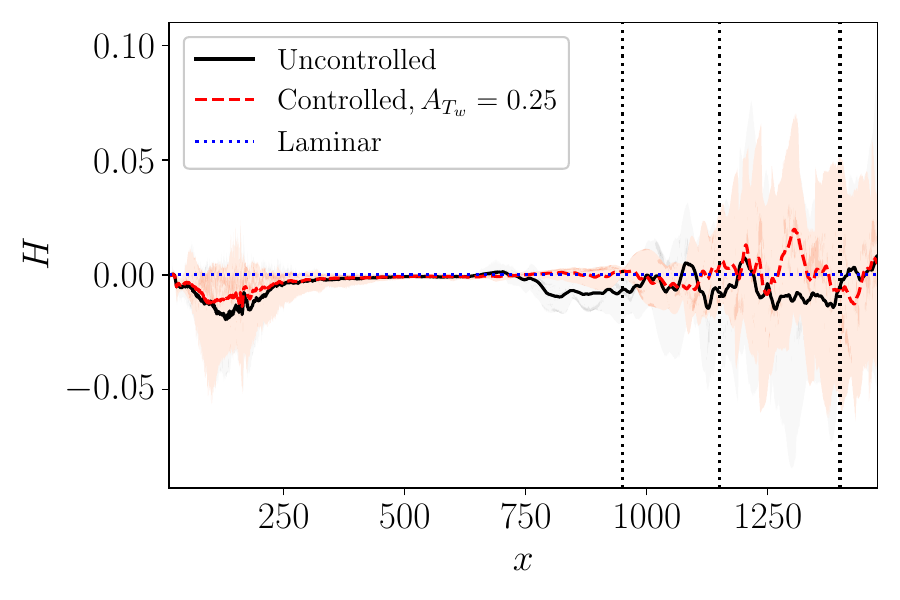}\label{fig:MDW_helicity}}
\hfill
  \centering 
  \subfloat[]{\includegraphics[width=0.8\textwidth]{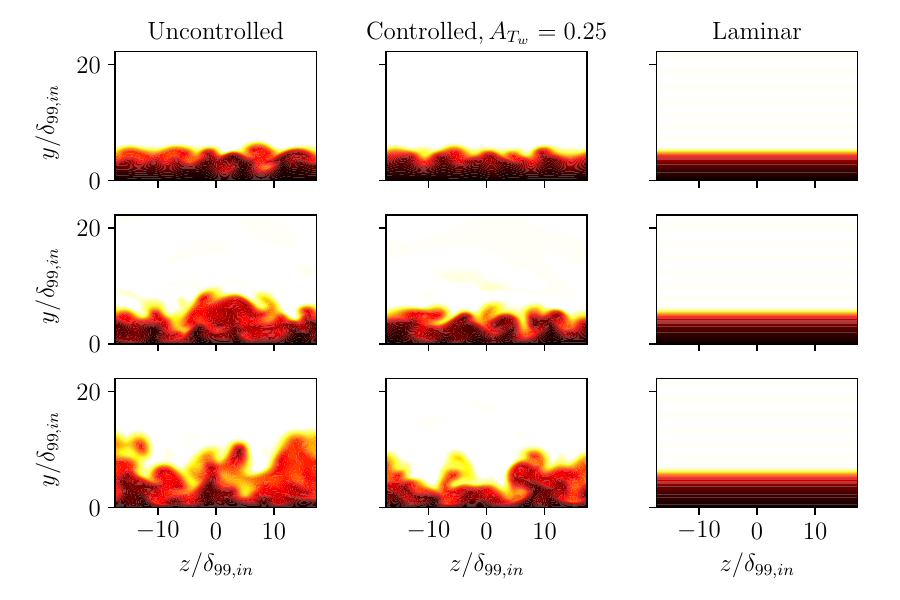}\label{fig:MD_contours}}
\caption{Effect of control method on (\textit{a}) time-average streamwise distribution of weighted helicity density (lines) with superimposed probability density function (shading), and (\textit{b}) instantaneous cross-sectional distribution of mass flux deficit at axial locations marked in (\textit{a}) by the black, dotted vertical lines; top to bottom: $x=950$, $1150$ and $1400$. Contour levels in (\textit{b}) go from zero (white) to one (dark red).}
\label{fig:SMF_symmetry}
\end{figure}

In section \ref{sec:transition_SMF}, it is shown that the heated streaks preserve the spanwise symmetry of the flow for longer streamwise distances.
This is further investigated following the streamwise evolution of the helicity, which provides a measure of structural coherence in the boundary layer. The helicity density ($h$) is integrated across the cross-section of the domain, and the quantity, $H$, is defined as 

\begin{equation}
\begin{cases}
	H = \frac{\iint (1-\rho u) \, h \, dydz}{\iint (1-\rho u) \, dydz}\\
	\hfill\\
	h = \frac{\overrightarrow{u} \cdot \nabla \times \overrightarrow{u}}{|\overrightarrow{u}||\nabla \times \overrightarrow{u}|} \,.
\end{cases}
\label{eq:helicity_MDW}
\end{equation}
In the preceding equation $H$ indicates the integral of the helicity density weighted with the mass flow deficit to isolate the contribution of the boundary layer. For a left-right symmetric flow $H=0$, whereas any deviation from zero indicates a loss of symmetry. The parameter $H$ is also computed for a fully laminar configuration for reference, and $H=0$ as the boundary layer is irrotational ($\nabla \times \overrightarrow{u}=0$). For both the uncontrolled and controlled configurations, the flow is asymmetric close to the start of the domain ($x<250$, Fig. \ref{fig:MDW_helicity}), due to the function $g(z)$ (equation \eqref{eq:actuator_eq_FL}), which breaks the symmetry of the forcing disturbance. Then both the configurations recover the symmetry until $x\approx750$, when the uncontrolled case has $H<0$ and the left-right symmetry is broken from there on as depicted from the cross-sectional distribution of the mass flow deficit in Fig. \ref{fig:MD_contours}. For the controlled configuration, fluctuations of $H$ become significant from $x\approx950$. This indicates that the control method preserves the symmetry for a longer streamwise distance, which is approximately $140\delta_{99,in}$. For $x \ge 1150$, the amplitude of the spatial oscillations for $H$ rapidly increases for both configurations due to transition to turbulence and the boundary layer thickness also increases compared to the laminar configuration (Fig. \ref{fig:MD_contours}).

\section{Effect of spanwise phase between control and disturbance actuator on control method effectiveness}\label{app:phase_effect_SMF}
\begin{figure}
  \centering 
  \subfloat[]{  \includegraphics[width=0.48\textwidth]{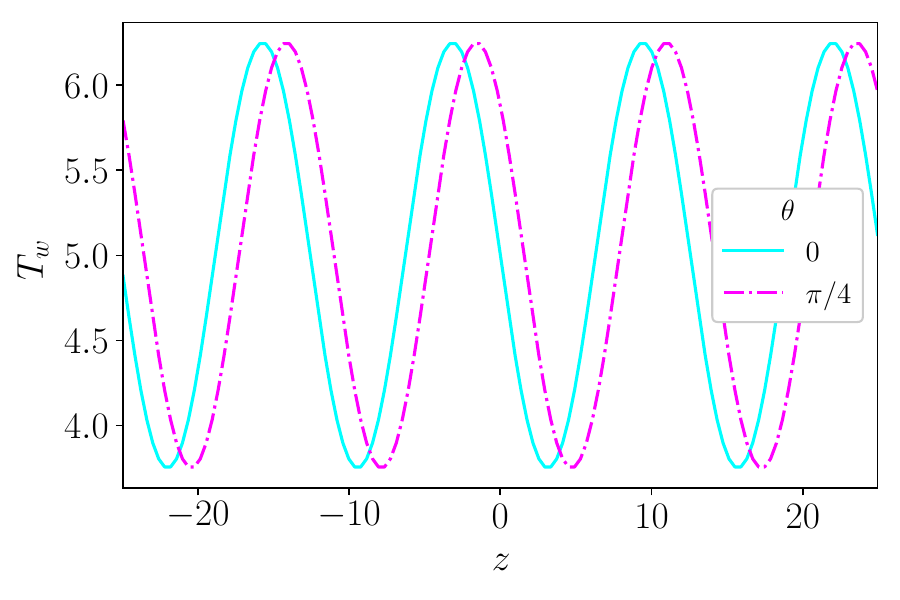}\label{fig:SMF_phi_effect_Tw_a}}
  \hfill
  \subfloat[]{\includegraphics[width=0.48\textwidth]{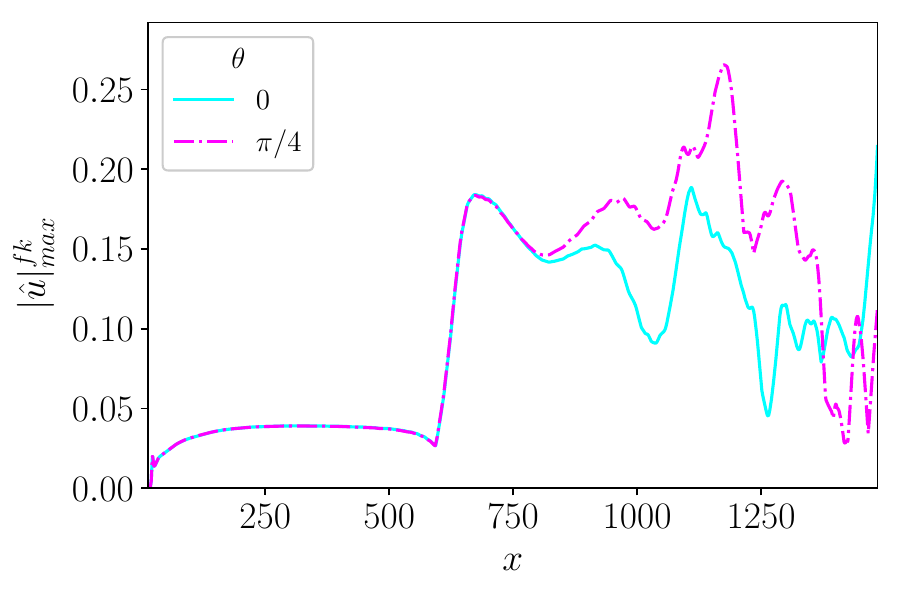}\label{fig:SMF_phi_effect_Fourier_streaks}}
\hfill
  \subfloat[]{\includegraphics[width=0.48\textwidth]{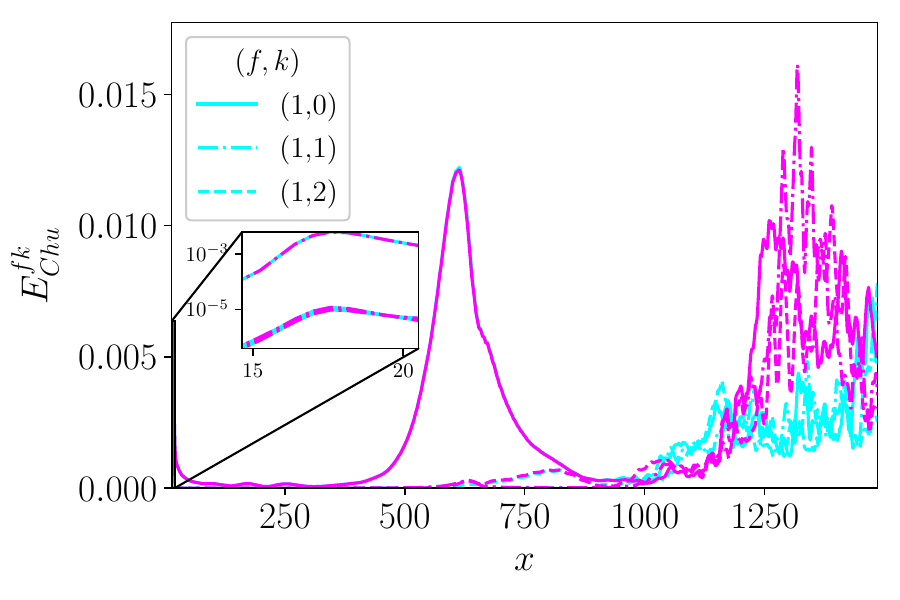}\label{fig:SMF_phi_effect_Fourier_Echu}}
\hfill
  \subfloat[]{\includegraphics[width=0.48\textwidth]{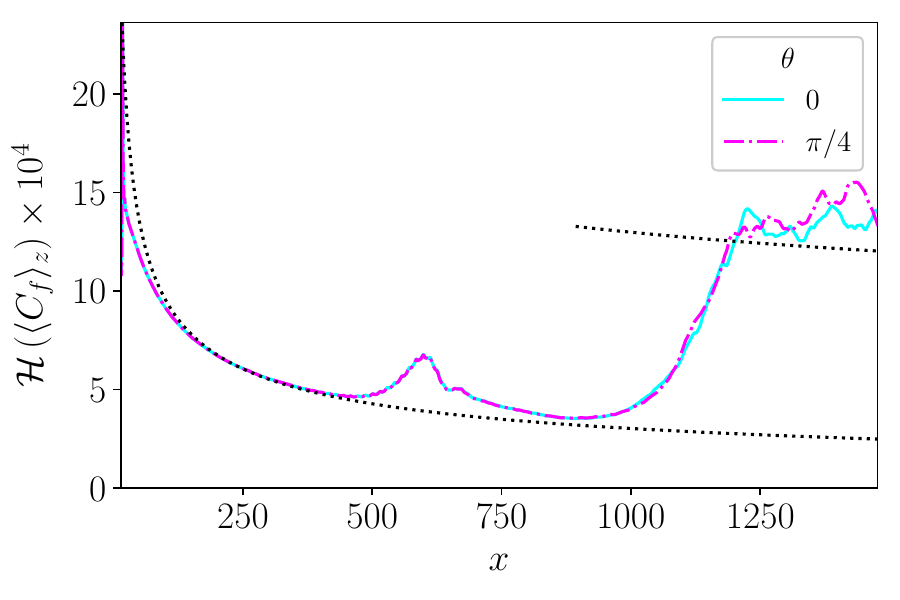}\label{fig:SMF_phi_effect_Cf}}    
\caption{Effect of spanwise phase ($\theta$) between the control law and the disturbance actuator on (\textit{a}) the spanwise wall temperature distributions, and on the  streamwise distribution of (\textit{b}) control-streaks amplitude, (\textit{c}) planar and oblique second Mack mode energy, and (\textit{d}) envelope of the spanwise-averaged skin friction coefficient. In (\textit{d}) grey dotted lines indicate the laminar and turbulent correlations.}
\label{fig:SMF_phi_effect_Tw}
\end{figure}

The results in section \ref{sec:transition_SMF} showed that the control method delays transition via a stabilization of the second Mack mode, and preserving the symmetries of the flow for a longer streamwise extent. It is shown that this is a result of constructive interference between the control-streaks and the streaks generated by triadic (difference) interactions. In this section, it is verified that this constructive interference is not the result of a mathematical construct due to the zero relative spanwise phase ($\theta$ in equation \eqref{eq:Tw_eq}) between the control law and the forcing disturbance (equation \eqref{eq:actuator_eq_FL}). Thus, the effectiveness of the method is here investigated for a configuration with $\theta=\pi/4$ and compared to the baseline configuration with $\theta=0$ (Fig. \ref{fig:SMF_phi_effect_Tw}). For convenience, the configurations with $\theta =0$ and $\theta=\pi/4$ are also referred to as in-phase and mistuned configuration in this section.

The effect of spanwise phase is evaluated for the configuration with $A_{T_w}=0.25$ presented in section \ref{sec:transition_SMF}, and the amplitude of the control streaks is the same for the in-phase and mistuned configurations ($x<500$ in Fig. \ref{fig:SMF_phi_effect_Fourier_streaks}). As a result, the energy of the second Mack mode planar, $(f,k)=(1,0)$, and oblique, $(f,k)=(1,[1,2])$, waves is unaffected by $\theta$ ($300<x<900$ in Fig. \ref{fig:SMF_phi_effect_Fourier_Echu}). The rapid growth of the amplitude of the control-streaks due to the interaction with the non-linear streaks that undergo transient growth is also unaffected by mistuning ($500<x<750$ in Fig. \ref{fig:SMF_phi_effect_Fourier_streaks}). For $x>900$, both the amplitude of the streaks and the second Mack mode energy are significantly affected by $\theta$. However, for $x>900$ there is onset of laminar to turbulence transition (Fig. \ref{fig:SMF_phi_effect_Cf}) and this is unaffected by mistuning between control and disturbance forcing. Overall, this indicates that the effectiveness of the control method is unaffected by the relative phase between the control streaks and the disturbance forcing, and therefore the placement of the hot and cold patches that generate the spanwise non-uniform surface temperature. This is a further indication that for this configuration the modification of the mean flow deformation due to the streaks dominates the stabilization of the second Mack mode, and the effect of the three dimensional, steady deformation of the boundary layer is negligible.

\bibliography{lb_icl}

\end{document}